\documentclass[preprint,letter,apjfonts]{emulateapj}
\usepackage{epsf}
\usepackage{apjfonts}
\usepackage{graphicx}

\newcommand{\oii}{[O\,{\sc ii}]}
\newcommand{\oiii}{[O\,{\sc iii}]}

\newcommand{\nii}{[N\,{\sc ii}]}

\newcommand{\sii}{[S\,{\sc ii}]}

\newcommand{\hei}{He\,{\sc i}}

\newcommand{\neiii}{[Ne\,{\sc iii}]}
\newcommand{\neiv}{[Ne\,{\sc iv}]}

\newcommand{\ariii}{[Ar\,{\sc iii}]}
\newcommand{\ariv}{[Ar\,{\sc iv}]}

\newcommand{\cliii}{[Cl\,{\sc iii}]}

\newcommand{\ha}{H$\alpha$}
\newcommand{\hb}{H$\beta$}

\newcommand{\kms}{km s$^{-1}$}

\received{}
\revised{}
\accepted{}

\slugcomment{}
\shorttitle{Dust and chemical abundances of the Sgr dwarf galaxy PN Hen2-436}
\shortauthors{Otsuka et al.}

\begin{document}

\title{Dust and Chemical Abundances of the Sagittarius dwarf Galaxy Planetary Nebula Hen2-436}

\author{Masaaki Otsuka\altaffilmark{1}
, Margaret Meixner\altaffilmark{1,2}
, David Riebel\altaffilmark{3}
, Siek Hyung\altaffilmark{4}
, Akito Tajitsu\altaffilmark{5}
, Hideyuki Izumiura\altaffilmark{6}
}
\altaffiltext{1}{Space Telescope Science
Institute, 3700 San Martin Drive, Baltimore, MD 21218, USA; otsuka@stsci.edu}
\altaffiltext{2}{Radio \& Geoastronomy Division, Harvard-Smithsonian for Astrophysics, 
60 Garden St. MS 42 Cambridge, MA 02138-1516, USA}
\altaffiltext{3}{Department of Physics and Astronomy, The Johns Hopkins University, 3400 North Charles St.
Baltimore, MD 21218, USA}
\altaffiltext{4}{School of Science Education (Astronomy), Chungbuk National
University, 12 Gaeshin-dong Heungduk-gu, CheongJu, Chungbuk 361-763, Korea}
\altaffiltext{5}{Subaru Telescope, NAOJ, 650 North A'ohoku Place, Hilo, HI 96720, USA}
\altaffiltext{6}{Okayama Astrophysical Observatory (OAO), NAOJ, 
Kamogata, Okayama 719-0232, Japan}

\begin{abstract}
We have estimated elemental abundances of the planetary nebula Hen2-436 in 
the Sagittarius (Sgr) spheroidal dwarf galaxy using ESO/VLT FORS2, Magellan/MMIRS, 
and $Spitzer$/IRS spectra. We have detected candidates of fluorine [F\,{\sc
ii}]$\lambda$4790, krypton [Kr\,{\sc iii}]$\lambda$6826, and phosphorus [P\,{\sc
ii}]$\lambda$7875 lines and successfully estimated the abundances of these elements 
([F/H]=+1.23, [Kr/H]=+0.26, [P/H]=+0.26) for the first time. These elements 
are known to be synthesized by neutron capture process in the He-rich intershell during the
thermally pulsing AGB phase. We present a relation between 
C, F, P, and Kr abundances among PNe and C-rich stars. The detections of 
F and Kr in Hen2-436 support the idea that F and Kr together with C 
are synthesized in the same layer and brought to the surface by the third dredge-up. We have detected 
N~{\sc ii} and O~{\sc ii} optical recombination lines (ORLs) and derived 
the N$^{2+}$ and O$^{2+}$ abundances. 
The discrepancy between the abundance derived from the oxygen ORL and that derived 
from the collisionally excited line is $>$1 dex. 
To investigate the status of the central star of the PN, nebula condition, 
and dust properties, we construct a theoretical spectral energy distribution (SED) model to match 
the observed SED with {\sc Cloudy}. By comparing the derived luminosity and temperature of 
the central star with theoretical evolutionary tracks, we conclude that the initial mass of 
the progenitor is likely to be $\sim$1.5-2.0 $M_{\odot}$ and the age is $\sim$3000 yr after the 
AGB phase. The observed elemental abundances of 
Hen2-436 can be explained by a theoretical nucleosynthesis model with a star of initial mass
2.25 $M_{\odot}$, $Z$=0.008 and LMC compositions. 
We have estimated the dust mass to be 2.9$\times$10$^{-4}$ $M_{\odot}$ (amorphous carbon 
only) or 4.0$\times$10$^{-4}$ $M_{\odot}$ (amorphous carbon and PAH). Based on the assumption that 
most of the observed dust is formed during the last two thermal pulses and the dust-to-gas mass 
ratio is 5.58$\times$10$^{-3}$, the dust mass-loss rate and the total mass-loss rate 
are $<$3.1$\times$10$^{-8}$ $M_{\odot}$ yr$^{-1}$ 
and $<$5.5$\times$10$^{-6}$ $M_{\odot}$ yr$^{-1}$, respectively. 
Our estimated dust mass-loss rate is comparable to a Sgr dwarf galaxy AGB star with similar metallicity and luminosity.
\end{abstract}

\section{Introduction}

Currently, $>$5,000 objects are regarded as planetary nebulae (PNe) in the local group
galaxies. Of them, BoBn1 (Otsuka et al. 2008; Otsuka
et al. 2010), Wray16-423, StWr2-21 (Kniazev et al. 2008; Zijlstra et al. 2006), and Hen2-436 (Walsh et al. 1997; Dudziak et al. 2000; 
this paper) belong to the Sagittarius (Sgr) dwarf galaxy. Sgr dwarf galaxy PNe are interesting objects as they 
provide direct insight into old, low-mass stars and also information on the history of the Galactic halo. 
In some galactic evolution scenarios, the galactic halo is partly built up from the tidal destruction and assimilation of dwarf galaxies. 
Sgr dwarf galaxy PNe are ideal laboratories and appropriate references to study the evolution of metal-deficient stars found in the Galactic halo and low-mass 
stars ($\lesssim$3.5 $M_{\odot}$) in the LMC. The metallicity of Sgr dwarf galaxy PNe is relatively low, e.g., 
$<$0.01 $Z_{\odot}$ in BoBn1 or $\sim$0.5 $Z_{\odot}$ in the others, which is equal to the typical metallicity of the LMC, 
and all the Sgr dwarf galaxy PNe are C-rich ([C/O]$\gtrsim0$; e.g., Zijlstra et al. 2006) based on gas-phased C and O abundances. 
The Sgr dwarf galaxy PNe and LMC PNe would complement each other in a study of metal-deficient PNe.
In addition, since the distance to the Sgr dwarf galaxy is well determined (24.8 kpc; Kunder \& Chaboyar 2009) 
and the interstellar reddening to the Sgr dwarf galaxy PNe is relatively low ($E(B-V)$ $\lesssim$0.1), we can accurately estimate intrinsic flux densities. 
Moreover, by using the spatially highly resolved images taken by Hubble Space Telescope ($HST$) and 8-m class ground based telescopes, we can estimate the size and shape 
of the nebulae. The nebular size is an important parameter in building spectral energy distributions (SED). 
We have collected data on several Sgr dwarf galaxy PNe to investigate elemental abundances, dust mass, and 
evolutionary status of the central star in a metal-deficient environment. 
In this paper, we analyze the most recently secured spectral data of the Sgr dwarf galaxy PN Hen2-436 (PN G004.8--22.7).

Hen2-436 is an interesting object in terms of chemical abundances and dust production in metal-poor 
environments. Zijlstra et al. (2006) argued that dust exists within the nebula. 
Sterling et al. (2009) detected [Kr\,{\sc iii}]$\lambda$2.19 $\mu$m and [Se~{\sc iv}]$\lambda$2.29 $\mu$m 
lines in the Gemini/GNIR spectra, although the amounts of these elements are not estimated yet. 
In the extra-Galactic PNe, these slow neutron capture elements ({\it s}-process elements) have so far 
only been detected in BoBn1 (Otsuka et al. 2010) and this nebula. The detection of such rare elements in PNe is highly interesting.

For Hen2-436, we performed a comprehensive 
chemical abundance analysis based on optical ESO/VLT FORS2 spectra, near-IR spectra from Magellan/MMIRS, and 
mid-IR $Spitzer$/IRS spectra. From FORS2 spectra, we found candidate detections of fluorine (F), phosphorus (P), and krypton (Kr) forbidden lines, 
and we estimate abundances of these elements. The estimations of F and P are done for the first time. Both elements are synthesized 
by neutron capture in the He-rich intershell of AGB stars. Through spectral energy distribution (SED) 
modeling with the photo-ionization code {\sc Cloudy} (Ferland 2004), 
we infer the evolutionary status of the central star and try to estimate the dust mass. 
We further investigate  the evolutionary status of the progenitors of Hen2-436 and the other Sgr dwarf galaxy PNe with similar metallicity. 
The observed chemical abundances are compared with theoretical nucleosynthesis model predictions.

\section{Data and Reduction} 

\subsection{VLT/FORS2 Archive Data}
The optical low-dispersion spectra of Hen2-436 
in the range 3320 {\AA} to 8620 {\AA} 
are available from the European Southern Observatory (ESO) archive.
The observations were performed by M.P\~{e}na (Prop. I.D.:
077.B-0430B) on August 23th, 2006, using 
the visual and near UV FOcal Reducer and low dispersion Spectrograph 2 (FORS2; Appenzeller et al. 1998)
at the Cassegrain focus of the ANTU, one of the four 8.2-m telescopes
of the ESO Very Large Telescope (VLT) at Paranal, Chile. The entrance slit 
size was 417$\farcs79$ in length and 1$\farcs01$
in width. The position
angle (P.A.) was set to 0$^{\circ}$. The 2$\times$2 on-chip binning pattern was chosen, hence 
the sampling pitch was $\sim$1.4 {\AA} pixel$^{-1}$ in wavelength 
and 0$\farcs5$ pixel$^{-1}$ in space. By measuring the FWHM ($\sim$4.7~\AA) of sky lines around 5500~\AA, we determine 
the spectral resolving power $R$ ($\lambda$/$\Delta\lambda$) to be $\sim$1200.
The data were taken by 
long (3$\times$60 sec) and short (3$\times$ 1, 5, and 10 sec) 
exposures. In order to measure the fluxes for strong lines such 
as {\oiii} and H$\alpha$, we used the short exposure frames, and for
weak line measurements we used the long exposure frames.  
The seeing was $\sim$1$\farcs0$ during the exposure. 
The standard star LTT7987 was observed for flux
calibration and telluric absorption correction. 
For wavelength calibration, we used He-Hg-Cd lamp frames 
and night sky-lines recorded in the object frames. 

Data reduction and emission line analysis were performed mainly with 
a long-slit reduction package noao.twodspec in 
IRAF\footnote[7]{IRAF is distributed by
the National Optical Astronomy Observatories, which are operated by 
the Association of Universities for Research in Astronomy (AURA), 
Inc., under a cooperative agreement with the National Science Foundation.}.
Data reduction was performed in a standard manner. In measuring the fluxes of 
emission-lines, we assumed that the line profiles 
were all Gaussian and we applied multiple Gaussian fitting techniques.

In the panels (a), (b), (c), and (d) of Figure \ref{spec}, we present the FORS2 spectra. 
The vertical axis is the scaled flux density in units of {\AA}$^{-1}$, normalized 
to the {\hb} flux $F$({\hb})=1000, and the horizontal axis is the rest wavelength. 
We also present candidate detections of fluorine [F\,{\sc ii}]$\lambda$4790, krypton 
[Kr\,{\sc iii}]$\lambda$6826, and phosphorus [P\,{\sc ii}]$\lambda$7875 lines.

\subsection{Magellan/MMIRS Observation}
We performed $J$ band (1.17-1.33 $\mu$m) moderate-resolution spectroscopy using the MMT and Magellan 
Infrared Spectrograph (MMIRS; McLeod et al. 2004) at the Cassegrain focus of 
the Magellan Clay 6.5-m telescope on April 27th, 2010 (PI.: M.Meixner). 
The detector of MMIRS is a 2048$\times$2048 pixel Hg-Cd-Te Hawaii-II array. The entrance 
slit size was 420$''$ in length and 1$''$ in width. 
The P.A. was set to 0$^{\circ}$. The sampling pitch is 
$\sim$2.5$\times$10$^{-4}$ $\mu$m pixel$^{-1}$ in wavelength 
and 0$\farcs2$ pixel$^{-1}$ in space.  We determine $R$ to be $\sim$1400 by measuring the FWHM 
($\sim$8.5$\times$10$^{-4}$ $\mu$m) of night sky-lines around 1.23 $\mu$m. We observed Hen2-436 
and the standard star HIP94663 (A0V, 2MASS $J$=7.425) using a three-point dither pattern for 
sky background subtraction. The total exposure time for Hen2-436 is 900 sec (3$\times$300 sec). 
The seeing was $\sim$1$''$ during the exposure. For wavelength calibration, 
we used night sky-lines recorded in the object frames. Data reduction was performed using IRAF. 
The telluric absorption and flux calibration were corrected using the HIP94663 frames. 
The MMIRS spectrum s presented in Figure \ref{spec}(e). 
The strong emission lines are He~{\sc i} + Pa$\beta$ $\lambda$1.28$\mu$m.

\subsection{Spitzer/IRS Archive Data}
\begin{figure*}
\epsscale{1.0}
\plotone{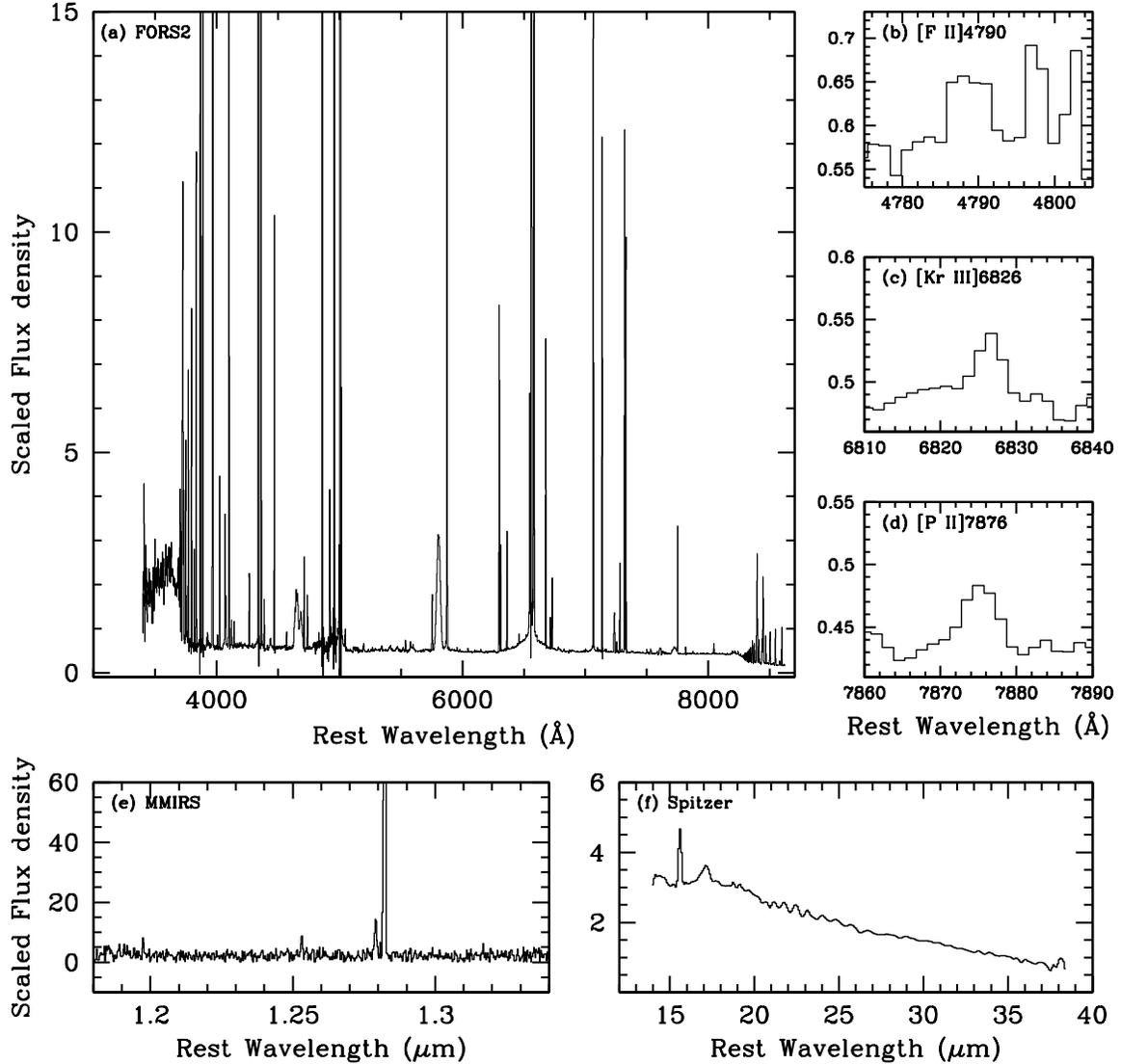}
\caption{({\it panels a, b, c, and d}) the FORS2 spectrum of Hen2-436. 
The vertical axis is the scaled flux density in unit of {\AA}$^{-1}$ which is normalized to the H$\beta$ flux $F$(H$\beta$)=1000. 
({\it panels e and f}) the MMIRS and $Spitzer$ spectrum of Hen2-436. The vertical axis is the scaled flux density in units of $\mu$m$^{-1}$ which 
is normalized to the $F$(H$\beta$)=1. \label{spec}}
\end{figure*}

We used the data set of program ID: P30333 (PI: A.Zijlstra)
taken by the Spitzer space telescope on October 18th, 2006. 
The data were taken with the Infrared Spectrograph (IRS, Houck et al. 2004) using the 
SL (5.2-14.5 $\mu$m) and LL (14-38$\mu$m) modules. Due to the low signal-to-noise ratio, the SL 
data were not available for chemical abundance analysis and SED modeling. The
one-dimensional spectra were extracted using {\it spice} version c15.0A. 
We extracted the region within $\pm$1$''$ from the center of each spectral
order and summed along the spatial direction. No correction 
for interstellar extinction was made since it is negligibly small in this IR wavelength band. 
In Figure \ref{spec}(f) we present the extracted Spitzer spectrum. 
We detected the [Ne\,{\sc iii}] $\lambda$15.5$\mu$m fine-structure line. 
The spectrum shows a broad emission feature to the red of the [Ne~{\sc iii}] $\lambda$15.5 $\mu$m line.
Since Hen2-436 is considered to be a C-rich PN, the feature might be related to the C-C-C PAH bending modes, which are observed in some 
Magellanic Cloud PNe (Bernard-Salas et al. 2009). It should be noted, however, that to date, the C and O abundances of 
Hen2-436 have only been derived from different types of emission lines; the C and O abundances are from 
recombination and from forbidden lines, respectively. There are no UV spectra available from which 
to estimate the C$^{2+}$ abundances using the C~{\sc iii}] $\lambda\lambda$1906/09 lines. Cohen \& Barlow (2005) found 
that the PAH 7.7 $\mu$m band is detectable in PNe with C/O $>$ 0.56$_{-0.41}^{+0.21}$ based on ISO/SWS spectra of over 40 PNe.
To verify whether Hen2-436 has PAHs, we need to check the C/O ratio based on the C and O abundances from 
the same type of emission lines.

\subsection{$HST$/WFPC2 Archive Data}
\begin{figure}
\epsscale{1.0}
\plotone{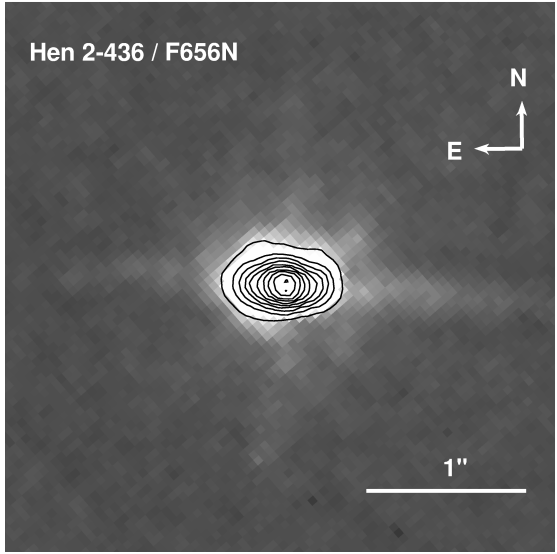}
\caption{The intensity contour map on the $HST$/WFPC2 F656N (H$\alpha$) image. \label{image}}
\end{figure}

The archival $HST$/WFPC2 images taken with the F502N ($\lambda_{c}$=5013 {\AA}/$\Delta\lambda$=35.8 {\AA}), 
F547M (5484 {\AA}/638 {\AA}), and F656N (6564 {\AA}/28 {\AA}) filters 
are publicly available. These images were taken by A.Zijlstra (Proposal I.D.: 9356) on May 4th, 2003. 
 The WFPC2 frames reduced and calibrated by the $HST$ pipeline 
(including MultiDrizzle), were downloaded from the $HST$ archive at the 
Canadian Astronomy Data Centre (CADC). 
In Figure \ref{image} we present the intensity contour map overlaid on top of the F656N image.
Hen2-436 has an elongated nebula along P.A.=90$^{\circ}$.  
We performed aperture photometry to estimate the amount of light passing through the slit 
and to do flux correction (see below). We measured the total flux within a 1$''$ radius 
and subtracted the background from an annulus centered on the PN with inner and
outer radii of 1.1$''$ and 1.3$''$. We performed aperture corrections using point spread functions generated by the TinyTim 
software package\footnote[8]{http://www.stsci.edu/software/tinytim/tinytim.html}. 
For F547M, we assumed a 9000 K blackbody function as the incident SED.
The measured fluxes are listed in Table \ref{hst}. $X(-Y)$ means $X\times10^{-Y}$ hereafter. 
The uncertainty corresponds to the standard deviation of the background. 
To estimate the contributions from both the stellar and nebular 
continuum to F502N and F656N, we measured the total emission line flux in the F547M band 
using the FORS2 spectra. Since the total emission line flux in the F547M band is 2.34(--13) $\pm$ 1.85(--14) 
erg s$^{-1}$ cm$^{-2}$, about half of the total flux could originate from the continuum. Using the 
averaged continuum flux density in the F547M,
we estimated the stellar and nebular continuum subtracted 
[O~{\sc iii}] and H$\alpha$ fluxes to be 5.29(--12) and 2.39(--12) erg s$^{-1}$ cm$^{-2}$, respectively.
 
The observation logs for Hen2-436 are summarized in Table \ref{obs_log}.

\begin{table}
\centering
\caption{Observation logs for Hen2-436 \label{obs_log}}
\begin{tabular}{cccl}
\hline\hline
Obs. Date&Telescopes/Instruments&Wavelength or&Exp.Time\\
         &         &Filter    &        \\
\hline
2006/08/23&VLT/FORS2&3320--8620 {\AA}&3$\times$60,10,5,1 sec\\
2010/04/27&Magellan/MMIRS&1.17-1.33 $\mu$m&3$\times$300 sec\\
2006/10/18&Spitzer/IRS&14-38 $\mu$m&2$\times$14.68 sec\\
2003/05/04&HST/WFPC2&F502N&2$\times$80 sec\\
          &         &F547M&1$\times$60 sec\\
          &         &F656N&2$\times$100 sec\\
\hline
\end{tabular}
\end{table}

\begin{table}
\centering
\caption{The fluxes measured from the $HST$ images. \label{hst}}
\begin{tabular}{ccc}
\hline\hline
Filter&$\lambda_{c}$&Flux\\
      &({\AA})      &(erg s$^{-1}$ cm$^{-2}$)\\  
\hline
F502N&5013&5.38(--12) $\pm$ 7.96(--14)\\
F547M&5484&5.69(--13) $\pm$ 6.06(--14)\\
F656N&6564&2.43(--12) $\pm$ 3.04(--14)\\
\hline
\end{tabular}
\end{table}

\section{Results} 
\subsection{Detected Lines and Interstellar Reddening Correction \label{irc}}
We have detected over 100 lines, including useful lines to
investigate plasma diagnostics and estimate ionic abundances. 
We estimate ionic abundances using detected optical recombination lines
(ORLs) and collisionally excited lines (CELs). 

Before performing plasma diagnostics and chemical abundance analysis, 
we corrected our measured fluxes for interstellar reddening by determining the reddening coefficient at {\hb}, $c$({\hb}). 
We compared the observed line ratio of {\ha} to {\hb} to 
the theoretical ratio computed by Storey \& Hummer (1995) assuming a temperature $T_{\epsilon}$ = 10$^{4}$ K 
a density $n_{\epsilon}$ = 10$^{4}$ cm$^{-3}$ and a nebula optically thick to Ly-$\alpha$ (Case B of Baker \& Menzel 1938). 
From this we estimate $c$({\hb}) = 0.23 $\pm$ 0.05. From Seaton's (1979) relation $c$({\hb}) 
= 1.47$E(B-V)$ one obtains $E(B-V)$ = 0.16 $\pm$ 0.03, 
which agrees fairly well with the value (0.13) measured in the direction of Hen2-436 in the Galactic extinction map of Schlegel et al. (1998).

All of the line fluxes in optical to near-IR were de-reddened using the formula: 
\medskip

\begin{equation}
\label{redc}
\log_{10}\left[\frac{I(\lambda)}{I{\rm (H\beta)}}\right] = 
\log_{10}\left[\frac{F(\lambda)}{F{\rm (H\beta)}}\right] + c({\rm H\beta})f(\lambda),
\end{equation}
\medskip

\noindent
where $I(\lambda)$ and $F(\lambda)$ are the de-reddened and the observed fluxes at $\lambda$, respectively, 
and $f(\lambda)$ is the interstellar extinction at $\lambda$, computed by 
the reddening law of Cardelli et al. (1989) with $R_{V}$ = 3.1. 
Comparing the observed [O~{\sc iii}]$\lambda$5007, the H$\alpha$ fluxes from the FORS2 observation 
and the continuum subtracted F507N and F656N fluxes, we estimate 83$\%$ of the total flux from Hen2-436 to be passing 
through the $1.\hspace{-2pt }''0$ width slit entrance of FORS2. We estimate the total {\hb} flux to be 6.37(--13) 
$\pm$ 9.76(--15) erg s$^{-1}$ cm$^{-2}$. Our measured value is consistent with Cahn et al. 
(1992), $\log_{10}F(H\beta)$ = --12.02 $\pm$ 0.03. Using $c$({\hb}), we estimate the de-reddened total {\hb} flux to be 1.08(--12) $\pm$ 1.46(--13) erg s$^{-1}$ cm$^{-2}$.

The de-reddened emission line fluxes detected in the FORS2, MMIRS, and IRS spectra are listed in Table \ref{list}. 
These line fluxes are normalized to $I$({\hb})=100. In the columns of laboratory wavelength ($\lambda_{\rm lab}$), 
we indicate the stellar origin lines by the asterisks ($\ast$). These stellar lines have the wide FWHM, typically, $\gtrsim$ 8 {\AA}.
Thanks to the large photon collecting power of VLT/FORS2, we detected 192 emission lines in total, including many weak intensity 
recombination lines of C, N, and O. Our measurements largely improved over previous detections 
by Welsh et al. (1997), who detected 51 emission lines in ESO NTT 3.5-m/EMMI (spectral resolution 4.5 {\AA}) 
and ESO 1.5-m/Boller and Chievens spectra (5-7 {\AA}). 
The normalized line fluxes $I$($\lambda$) of Welsh et al. (1997) and our measurements agree within 2 $\%$.

\subsection{Plasma Diagnostics}
\subsubsection{CELs Plasma Diagnostics}
\begin{figure}
\epsscale{1.1}
\plotone{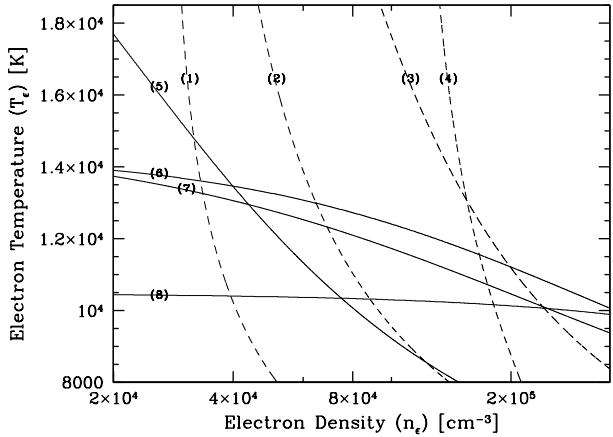}
\caption{Plasma diagnostic diagram. Each curve labeled with an ID is calculated by the following 
line intensity ratios; (1) [S\,{\sc ii}] ($\lambda\lambda$6716/31)/($\lambda\lambda$4069/76), (2) 
[O\,{\sc ii}] ($\lambda$3727)/($\lambda\lambda$7320/30), (3) [Ar\,{\sc iv}] ($\lambda$4740)
/($\lambda$7170+$\lambda$7263), (4) [Cl\,{\sc iii}] ($\lambda$5517)/($\lambda$5537), 
(5) [N\,{\sc ii}] ($\lambda\lambda$6548/83)/($\lambda$5755), 
(6) [O\,{\sc iii}] ($\lambda$4959+$\lambda$5007)/($\lambda$4363),
(7) [Ne\,{\sc iii}] ($\lambda$15.5$\mu$m)/($\lambda$3869), and 
(8) [Ar\,{\sc iii}] ($\lambda$7135+$\lambda$7751)/($\lambda$5192). 
The solid lines indicate diagnostic lines of the electron temperature. The broken lines indicate diagnostic lines of the 
electron density. For $T_{\epsilon}$($[$N\,{\sc ii}$]$) and $n_{\epsilon}$($[$O\,{\sc ii}$]$),
 we corrected for recombination contributions to 
$[$N\,{\sc ii}$]$ $\lambda$5755 and $[$O\,{\sc ii}$]$$\lambda\lambda$7320/30, respectively.
The estimated $n_{\epsilon}$ and $T_{\epsilon}$ are summarized in Table \ref{diano_table}.
}
\label{diagno_figure}
\end{figure}

We have detected several collisionally excited lines (CEL) useful 
for estimations of the electron temperatures ($T_{\epsilon}$) and densities 
($n_{\epsilon}$).  
We have examined the electron temperature and density
structure within the nebula using 8 diagnostic CEL ratios. Electron
temperature and density were derived from each diagnostic ratio 
for each line combination by solving for the level populations in a multi-level 
($>$ 5 for most ions) atomic model using the collision strengths $\Omega_{ij}$ ($j$$>$$i$) 
and spontaneous transition probabilities $A_{ji}$ for each ion. 
The estimated electron temperatures and densities are
listed in Table \ref{diano_table}. We used the same atomic data as in Otsuka et al. (2010).
A diagnostic diagram that plots the loci 
of the observed diagnostic line ratios on the log($n_{\epsilon}$)-$T_{\epsilon}$ 
plane is shown in Figure \ref{diagno_figure}. The solid lines indicate diagnostics of the 
electron temperature, and the broken lines are electron density diagnostics. This diagram shows that
most of the CELs having ionization potentials (I.P.) $>$ 13.5 eV are emitted from
$T_{\epsilon}\sim$10\,000--14\,000 K and $n_{\epsilon}\sim$80\,000--200\,000 cm$^{-3}$ ionized gas. 
In such a high density nebula, {\sii} and {\oii} diagnostic line
ratios do not give reliable $T_{\epsilon}$ due to the
low critical density of {\sii}$\lambda\lambda$6716/31 and {\oii}$\lambda$3727 lines. 
Hence, these diagnostic lines were used only for density derivation. Meanwhile, the
{\nii}, {\oiii}, {\neiii}, and {\ariii} diagnostic lines can be used as
reliable $T_{\epsilon}$ indicators. 

For {\nii}$\lambda$5755 and
{\oii}$\lambda\lambda$7320/30, we need to consider the recombination contamination from N$^{2+}$ and O$^{2+}$ lines. 
We subtracted this contamination using Equations (\ref{rni}) and (\ref{roii}) given by Liu et al. (2000), 
\medskip

\begin{equation}
\label{rni}
\frac{I_{R}(\rm [N\,{\sc II}]\lambda5755)}{I(\rm H\beta)} = 
3.19\left(\frac{T_{\epsilon}}{10^4}\right)^{0.33}\times\frac{\rm N^{2+}}{\rm H^{+}},
\end{equation}

\begin{equation}
\label{roii}
\frac{I_{R}(\rm [O\,{\sc II}]\lambda\lambda7320/30)}{I(\rm H\beta)} = 
9.36\left(\frac{T_{\epsilon}}{10^4}\right)^{0.44}\times\frac{\rm O^{2+}}{\rm H^{+}}.
\end{equation}
\medskip

\noindent
where N$^{2+}$/H$^{+}$ and O$^{2+}$/H$^{+}$ are the doubly ionized
nitrogen and oxygen abundances, respectively. Adopting the value derived
from optical recombination line (ORL) analysis (see Section \ref{ir}), we estimate $I_{R}$({\rm
{\nii}} $\lambda$5755) $\sim$0.2, which is approximately 22 $\%$ of the
observed value. Likewise, we estimated $I_{R}$({\rm {\oii}}
$\lambda\lambda$7320/30) $\gtrsim$2.9, which is approximately $\gtrsim$25 $\%$ of the
observed value.

First, we calculated the electron density. We adopted a constant temperature of 
12\,000 K for the calculations of $n_{\epsilon}$({\sii}) and $n_{\epsilon}$({\oii}), and 14\,000 K when calculating $n_{\epsilon}$({\cliii}) and $n_{\epsilon}$({\ariv}). 
Next, we calculated the electron 
temperature. $T_{\epsilon}$({\nii}) was calculated assuming a density of $n_{\epsilon}$({\oii}), 
$T_{\epsilon}$({\ariii}) was using $n_{\epsilon}$({\cliii}),
and $T_{\epsilon}$({\neiii}) and $T_{\epsilon}$({\oiii}) were using 
$n_{\epsilon}$({\ariv}). The resultant $n_{\epsilon}$ 
and $T_{\epsilon}$ are listed in Table \ref{diano_table}.

\subsubsection{ORLs Plasma Diagnostics}
We detected optical recombination lines (ORLs) of He, C, N, and O. To
calculate ionic abundances of these elements using ORLs, we estimated the electron temperature from
the Balmer discontinuity and {\hei} line ratios. We used the same atomic data employed by Otsuka et al. (2010).

The electron temperature derived by these two methods are listed in Table \ref{diano_table}. The Balmer
discontinuity electron temperature $T_{\epsilon}$(BJ) of
14\,500 $\pm$ 2200 K was estimated using the equation given in  Liu et al. (2001). 

Using four diagnostic {\hei} line ratios, 
the {\hei} electron temperature $T_{\epsilon}$({\hei})  was estimated assuming a
constant electron density $n_{\epsilon}$ = 10$^{5}$ cm$^{-3}$. All the {\hei} lines we chose 
for the $T_{\epsilon}$({\hei}) derivation are insensitive to electron density.
We adopted the emissivities of {\hei} lines from Benjamin et al. (1999). 
We consider that the $T_{\epsilon}$({\hei}) from the 
$\lambda$7281/$\lambda$6678 ratio is the most reliable value 
because (i) {\hei} $\lambda$6678 and $\lambda$7281 are 
from levels having the same spin as the ground state and the
Case B recombination coefficients for these lines by Benjamin et
al. (1999) are more certain than those for $\lambda$4471 and
$\lambda$5876; (ii) the interstellar extinction is less of a complication due to the similar
wavelengths of these lines. Thus, we adopted $T_{\epsilon}$({\hei}) derived from the 
He\,{\sc i} $\lambda$7281/$\lambda$6678 ratio for the He$^{+}$ abundance calculation.

\begin{table*}
\centering
\scriptsize
\caption{The line list of Hen 2-436 detected in the FORS2/MMIRS/IRS 
observations. \label{list}}
\begin{tabular}{c@{\hspace{7pt}}l@{\hspace{5pt}}r@{\hspace{5pt}}r@{\hspace{5pt}}r|@{\hspace{4pt}}
@{}c@{\hspace{7pt}}l@{\hspace{5pt}}r@{\hspace{5pt}}r@{\hspace{5pt}}r|@{\hspace{4pt}}
@{}c@{\hspace{7pt}}l@{\hspace{5pt}}r@{\hspace{5pt}}r@{\hspace{5pt}}r}
\hline\hline
$\lambda_{\rm lab}$ &Ion&$f$($\lambda$)&$I$({\hb})&$\delta I$({\hb})&$\lambda_{\rm lab}$ &Ion&$f$($\lambda$)&$I$({\hb})&$\delta I$({\hb})&$\lambda_{\rm lab}$ &Ion&$f$($\lambda$)&$I$({\hb})&$\delta I$({\hb})\\
\hline
3498.77	&	 He~{\sc i} 	&	0.367	&	0.516	&	0.371	&	5131.25	&	 [Kr~{\sc v}]? 	&	--0.068	&	0.081	&	0.022	&	7099.80	&	 [Pb~{\sc ii}]? 	&	--0.369	&	0.075	&	 0.006 \\
3587.28	&	 He~{\sc i} 	&	0.349	&	0.659	&	0.180	&	5145.17	&	 C~{\sc ii} 	&	--0.071	&	0.074	&	0.017	&	7113.04	&	 C~{\sc ii} 	&	--0.371	&	0.053	&	 0.008 \\
3634.24	&	 He~{\sc i} 	&	0.341	&	0.668	&	0.102	&	5172.34	&	 N~{\sc ii} 	&	--0.077	&	0.015	&	0.008	&	7135.80	&	 [Ar~{\sc iii}] 	&	--0.374	&	7.664	&	 0.385 \\
3643.93	&	 Ne~{\sc ii} 	&	0.339	&	0.394	&	0.214	&	5179.90	&	 C~{\sc iii} 	&	--0.078	&	0.010	&	0.013	&	7160.61	&	 He~{\sc i} 	&	--0.377	&	0.078	&	 0.016 \\
3679.35	&	 H~{\sc i} 	&	0.332	&	0.123	&	0.110	&	5191.82	&	 [Ar~{\sc iii}] 	&	--0.081	&	0.063	&	0.005	&	7170.50	&	 [Ar~{\sc iv}] 	&	--0.378	&	0.035	&	 0.006 \\
3682.81	&	 He~{\sc i} 	&	0.331	&	0.182	&	0.108	&	5197.90	&	 [N~{\sc i}] 	&	--0.082	&	0.142	&	0.007	&	7231.34	&	 C~{\sc ii} 	&	--0.387	&	0.345	&	 0.039 \\
3686.83	&	 H~{\sc i} 	&	0.330	&	0.436	&	0.173	&	5259.06	&	 C~{\sc ii} 	&	--0.095	&	0.029	&	0.005	&	7236.42	&	 C~{\sc ii} 	&	--0.387	&	0.559	&	 0.036 \\
3691.55	&	 H~{\sc i} 	&	0.329	&	0.643	&	0.118	&	5261.19	&	 N~{\sc ii} 	&	--0.096	&	0.051	&	0.017	&	7254.15	&	 O~{\sc i} 	&	--0.390	&	0.182	&	 0.022 \\
3697.15	&	 H~{\sc i} 	&	0.328	&	0.713	&	0.151	&	5270.40	&	 [Fe~{\sc iii}] 	&	--0.098	&	0.079	&	0.011	&	7262.70	&	 [Ar~{\sc iv}] 	&	--0.391	&	0.043	&	 0.021 \\
3704.98	&	 He~{\sc i} 	&	0.327	&	1.786	&	0.134	&	5335.65	&	 [Fe~{\sc ii}] 	&	--0.111	&	0.042	&	0.004	&	7281.35	&	 He~{\sc i} 	&	--0.393	&	1.329	&	 0.064 \\
3711.97	&	 H~{\sc i} 	&	0.325	&	1.407	&	0.110	&	5344.31	&	 Fe~{\sc iii} 	&	--0.113	&	0.108	&	0.009	&	7298.04	&	 He~{\sc i} 	&	--0.395	&	0.064	&	 0.038 \\
3721.94	&	 H~{\sc i} 	&	0.323	&	2.041	&	0.158	&	5376.19$^{\ast}$	&	 C~{\sc iii} 	&	--0.119	&	0.797	&	0.052	&	7319.46	&	 [O~{\sc ii}] 	&	--0.398	&	8.293	&	 0.404 \\
3726.03	&	 [O~{\sc ii}] 	&	0.322	&	9.249	&	0.424	&	5403.40	&	 C~{\sc ii} 	&	--0.124	&	0.054	&	0.006	&	7330.20	&	 [O~{\sc ii}] 	&	--0.400	&	6.746	&	 0.354 \\
3734.37	&	 H~{\sc i} 	&	0.321	&	2.241	&	0.157	&	5426.70	&	 [Fe~{\sc iii}] 	&	--0.129	&	0.046	&	0.003	&	7468.31	&	 N~{\sc i} 	&	--0.418	&	0.016	&	 0.011 \\
3750.15	&	 H~{\sc i} 	&	0.317	&	3.066	&	0.164	&	5470.68	$^{\ast}$&	 C~{\sc iv} 	&	--0.136	&	0.066	&	0.006	&	7499.85	&	 He~{\sc i} 	&	--0.422	&	0.059	&	 0.005 \\
3759.88	&	 O~{\sc iii} 	&	0.315	&	0.351	&	0.118	&	5471.10$^{\ast}$	&	 C~{\sc iv} 	&	--0.137	&	0.097	&	0.009	&	7499.85	&	 He~{\sc i} 	&	--0.422	&	0.067	&	 0.010 \\
3770.63	&	 H~{\sc i} 	&	0.313	&	3.947	&	0.171	&	5480.05	&	 N~{\sc ii} 	&	--0.139	&	0.046	&	0.003	&	7530.80	&	 C~{\sc ii} 	&	--0.427	&	0.101	&	 0.012 \\
3784.89	&	 He~{\sc i} 	&	0.310	&	0.214	&	0.147	&	5488.46	&	 C~{\sc ii} 	&	--0.140	&	0.034	&	0.002	&	7725.90$^{\ast}$	&	 C~{\sc iv} 	&	--0.452	&	0.771	&	 0.094 \\
3797.90	&	 H~{\sc i} 	&	0.307	&	5.126	&	0.216	&	5512.77	&	 O~{\sc i} 	&	--0.144	&	0.045	&	0.005	&	7751.10	&	 [Ar~{\sc iii}] 	&	--0.455	&	1.842	&	 0.099 \\
3819.60	&	 He~{\sc i} 	&	0.302	&	1.554	&	0.451	&	5517.66	&	 [Cl~{\sc iii}] 	&	--0.145	&	0.057	&	0.007	&	7816.14	&	 He~{\sc i} 	&	--0.464	&	0.083	&	 0.007 \\
3835.38	&	 H~{\sc i} 	&	0.299	&	7.821	&	0.505	&	5537.89	&	 [Cl~{\sc iii}] 	&	--0.149	&	0.160	&	0.014	&	7816.14	&	 He~{\sc i} 	&	--0.464	&	0.083	&	 0.007 \\
3868.77	&	 [Ne~{\sc iii}] 	&	0.291	&	53.568	&	1.974	&	5543.47	&	 N~{\sc ii} 	&	--0.150	&	0.025	&	0.002	&	7875.99	&	 [P~{\sc ii}] 	&	--0.471	&	0.042	&	 0.008 \\
3889.05	&	 H~{\sc i} 	&	0.286	&	17.481	&	0.830	&	5552.68	&	 N~{\sc ii} 	&	--0.151	&	0.034	&	0.003	&	8046.30	&	 [Cl~{\sc iv}] 	&	--0.492	&	0.165	&	 0.012 \\
3920.68	&	 C~{\sc ii} 	&	0.279	&	0.175	&	0.055	&	5577.34	&	 [O~{\sc i}] 	&	--0.156	&	0.239	&	0.011	&	8196.50	&	 C~{\sc iii} 	&	--0.510	&	0.046	&	 0.006 \\
3926.54	&	 He~{\sc i} 	&	0.277	&	0.181	&	0.046	&	5602.44	&	 [K~{\sc vi}] 	&	--0.160	&	0.074	&	0.009	&	8203.93	&	 He~{\sc i} 	&	--0.511	&	0.062	&	 0.011 \\
3970.07	&	 H~{\sc i} 	&	0.266	&	30.116	&	1.037	&	5592.37$^{\ast}$	&	 O~{\sc iii} 	&	--0.164	&	0.032	&	0.008	&	8214.50	&	 C~{\sc iii} 	&	--0.513	&	0.051	&	 0.007 \\
4026.18	&	 He~{\sc i} 	&	0.251	&	2.564	&	0.212	&	5754.64	&	 [N~{\sc ii}] 	&	--0.185	&	0.918	&	0.120	&	8236.79$^{\ast}$	&	 He~{\sc ii} 	&	--0.515	&	0.027	&	 0.008 \\
4068.60	&	 [S~{\sc ii}] 	&	0.239	&	2.067	&	0.245	&	5794.88	&	 He~{\sc ii} 	&	--0.191	&	3.528	&	0.999	&	8240.19	&	 H~{\sc i} 	&	--0.516	&	0.044	&	 0.005 \\
4076.35	&	 [S~{\sc ii}] 	&	0.237	&	0.811	&	0.226	&	5811.97$^{\ast}$	&	 C~{\sc iv} 	&	--0.193	&	10.124	&	1.300	&	8260.93	&	 H~{\sc i} 	&	--0.518	&	0.019	&	 0.004 \\
4101.73	&	 H~{\sc i} 	&	0.229	&	23.668	&	0.726	&	5875.62	&	 He~{\sc i} 	&	--0.203	&	19.558	&	0.546	&	8264.62	&	 He~{\sc i} 	&	--0.518	&	0.033	&	 0.003 \\
4120.81	&	 He~{\sc i} 	&	0.224	&	0.479	&	0.282	&	5889.79	&	 C~{\sc ii} 	&	--0.205	&	0.030	&	0.002	&	8271.93	&	 H~{\sc i} 	&	--0.519	&	0.003	&	 0.001 \\
4143.76	&	 He~{\sc i} 	&	0.217	&	0.376	&	0.275	&	5896.78	&	 He~{\sc ii} 	&	--0.206	&	0.015	&	0.009	&	8276.31	&	 H~{\sc i} 	&	--0.520	&	0.000	&	 0.000 \\
4229.27	&	 [Fe~{\sc v}]? 	&	0.191	&	0.137	&	0.032	&	5910.58	&	 [Fe~{\sc ii}] 	&	--0.208	&	0.008	&	0.005	&	8281.12	&	 H~{\sc i} 	&	--0.520	&	0.019	&	 0.009 \\
4267.15	&	 C~{\sc ii} 	&	0.179	&	1.102	&	0.032	&	5927.82	&	 N~{\sc ii} 	&	--0.211	&	0.016	&	0.004	&	8286.43	&	 H~{\sc i} 	&	--0.521	&	0.029	&	 0.012 \\
4292.21	&	 O~{\sc ii} 	&	0.172	&	0.040	&	0.024	&	5931.83	&	 He~{\sc ii} 	&	--0.211	&	0.026	&	0.003	&	8292.31	&	 H~{\sc i} 	&	--0.521	&	0.043	&	 0.009 \\
4340.46	&	 H~{\sc i} 	&	0.156	&	46.149	&	1.177	&	5958.38	&	 O~{\sc i} 	&	--0.215	&	0.070	&	0.011	&	8298.83	&	 H~{\sc i} 	&	--0.522	&	0.070	&	 0.009 \\
4363.21	&	 [O~{\sc iii}] 	&	0.149	&	13.901	&	0.502	&	5977.03	&	 He~{\sc ii} 	&	--0.218	&	0.034	&	0.009	&	8305.90	&	 H~{\sc i} 	&	--0.523	&	0.103	&	 0.011 \\
4387.93	&	 He~{\sc i} 	&	0.141	&	0.671	&	0.039	&	5998.71	&	 [Cu~{\sc iii}] 	&	--0.221	&	0.027	&	0.007	&	8312.10	&	 C~{\sc iii} 	&	--0.524	&	0.125	&	 0.010 \\
4414.90	&	 O~{\sc ii} 	&	0.133	&	0.084	&	0.029	&	6046.23	&	 O~{\sc i} 	&	--0.228	&	0.056	&	0.004	&	8323.42	&	 H~{\sc i} 	&	--0.525	&	0.181	&	 0.014 \\
4439.71	&	 [Fe~{\sc ii}] 	&	0.125	&	0.333	&	0.028	&	6101.79	&	 [K~{\sc iv}] 	&	--0.235	&	0.031	&	0.005	&	8333.78	&	 H~{\sc i} 	&	--0.526	&	0.209	&	 0.015 \\
4471.47	&	 He~{\sc i} 	&	0.115	&	5.911	&	0.150	&	6118.26	&	 He~{\sc ii} 	&	--0.238	&	0.030	&	0.006	&	8345.55	&	 H~{\sc i} 	&	--0.527	&	0.249	&	 0.017 \\
4571.10	&	 Mg~{\sc i}] 	&	0.084	&	0.224	&	0.021	&	6141.70	&	 Ba~{\sc ii}? 	&	--0.241	&	0.018	&	0.006	&	8359.00	&	 H~{\sc i} 	&	--0.529	&	0.408	&	 0.027 \\
4634.12$^{\ast}$	&	 N~{\sc iii} 	&	0.065	&	0.877	&	0.040	&	6151.27	&	 C~{\sc ii} 	&	--0.242	&	0.050	&	0.008	&	8374.48	&	 He~{\sc i} 	&	--0.531	&	0.294	&	 0.019 \\
4640.64$^{\ast}$	&	 N~{\sc iii} 	&	0.063	&	0.713	&	0.034	&	6159.12	&	 [Mn~{\sc iii}] 	&	--0.244	&	0.018	&	0.010	&	8397.42	&	 He~{\sc i} 	&	--0.533	&	2.058	&	 0.139 \\
4650.25$^{\ast}$	&	 C~{\sc iii} 	&	0.060	&	1.679	&	0.050	&	6300.30	&	 [O~{\sc i}] 	&	--0.263	&	5.523	&	0.197	&	8413.32	&	 H~{\sc i} 	&	--0.535	&	0.328	&	 0.058 \\
4658.64$^{\ast}$	&	 C~{\sc iv} 	&	0.058	&	1.439	&	0.050	&	6312.10	&	 [S~{\sc iii}] 	&	--0.264	&	1.634	&	0.092	&	8421.67	&	 He~{\sc ii} 	&	--0.536	&	0.020	&	 0.003 \\
4665.86$^{\ast}$	&	 C~{\sc iii} 	&	0.056	&	0.948	&	0.046	&	6346.97	&	 Mg~{\sc ii} 	&	--0.269	&	0.034	&	0.007	&	8437.95	&	 H~{\sc i} 	&	--0.537	&	0.383	&	 0.047 \\
4673.73	&	 O~{\sc ii} 	&	0.053	&	0.522	&	0.032	&	6363.78	&	 [O~{\sc i}] 	&	--0.271	&	1.829	&	0.069	&	8444.55	&	 He~{\sc i} 	&	--0.538	&	1.430	&	 0.113 \\
4676.23	&	 O~{\sc ii} 	&	0.052	&	0.339	&	0.029	&	6461.71	&	 N~{\sc ii} 	&	--0.284	&	0.159	&	0.010	&	8467.25	&	 H~{\sc i} 	&	--0.540	&	0.428	&	 0.028 \\
4687.55	&	 [Fe~{\sc ii}] 	&	0.049	&	0.564	&	0.027	&	6548.04	&	 [N~{\sc ii}] 	&	--0.296	&	3.589	&	0.154	&	8480.79	&	 He~{\sc i} 	&	--0.542	&	0.023	&	 0.017 \\
4713.17	&	 He~{\sc i} 	&	0.042	&	1.379	&	0.028	&	6562.77	&	 H~{\sc i} 	&	--0.298	&	285.000	&	11.874	&	8486.29	&	 He~{\sc i} 	&	--0.543	&	0.024	&	 0.020 \\
4740.17	&	 [Ar~{\sc iv}] 	&	0.034	&	0.725	&	0.019	&	6583.46	&	 [N~{\sc ii}] 	&	--0.300	&	10.915	&	0.504	&	8500.32	&	 [Cl~{\sc iii}] 	&	--0.544	&	0.110	&	 0.029 \\
4756.59	&	 [Mn~{\sc iii}] 	&	0.029	&	0.043	&	0.031	&	6678.15	&	 He~{\sc i} 	&	--0.313	&	4.923	&	0.194	&	8502.48	&	 H~{\sc i} 	&	--0.544	&	0.428	&	 0.037 \\
4789.45	&	 [F~{\sc ii}] 	&	0.020	&	0.106	&	0.018	&	6716.44	&	 [S~{\sc ii}] 	&	--0.318	&	0.488	&	0.053	&	8519.35$^{\ast}$	&	 He~{\sc ii} 	&	--0.546	&	0.030	&	 0.013 \\
4802.58	&	 Ne~{\sc ii} 	&	0.016	&	0.088	&	0.020	&	6730.81	&	 [S~{\sc ii}] 	&	--0.320	&	1.098	&	0.058	&	8528.97	&	 He~{\sc i} 	&	--0.547	&	0.052	&	 0.014 \\
4861.33	&	 H~{\sc i} 	&	0.000	&	100.000	&	2.165	&	6779.94	&	 C~{\sc ii} 	&	--0.326	&	0.050	&	0.007	&	8545.38	&	 H~{\sc i} 	&	--0.549	&	0.544	&	 0.035 \\
4921.93	&	 He~{\sc i} 	&	--0.016	&	1.649	&	0.041	&	6791.47	&	 C~{\sc ii} 	&	--0.328	&	0.067	&	0.011	&	8566.92	&	 He~{\sc ii} 	&	--0.551	&	0.005	&	 0.006 \\
4931.23	&	 [O~{\sc iii}] 	&	--0.019	&	0.151	&	0.008	&	6800.68	&	 C~{\sc ii} 	&	--0.329	&	0.018	&	0.004	&	8581.88	&	 He~{\sc i} 	&	--0.552	&	0.117	&	 0.014 \\
4958.91	&	 [O~{\sc iii}] 	&	--0.026	&	265.073	&	6.368	&	6826.38	&	 [Kr~{\sc iii}] 	&	--0.333	&	0.036	&	0.005	&	8598.39	&	 H~{\sc i} 	&	--0.554	&	0.592	&	 0.039 \\
5006.84	&	 [O~{\sc iii}] 	&	--0.038	&	792.066	&	16.075	&	6907.80	&	 [Fe~{\sc iii}] 	&	--0.343	&	0.058	&	0.017	&	8616.95	&	 [Fe~{\sc ii}] 	&	--0.556	&	0.021	&	 0.012 \\
5041.26	&	 O~{\sc ii} 	&	--0.046	&	0.052	&	0.033	&	6933.89	&	 He~{\sc i} 	&	--0.347	&	0.016	&	0.007	&	 1.253$\mu$m 	&	 [Fe~{\sc ii}] 	&	--0.759	&	0.442	&	 0.133\\           
5047.74	&	 He~{\sc i} 	&	--0.048	&	0.262	&	0.018	&	6989.46	&	 He~{\sc i} 	&	--0.354	&	0.018	&	0.007	&	 1.279$\mu$m 	&	 He~{\sc i} 	&	--0.767	&	1.129	&	 0.212\\           
5114.06	&	 O~{\sc v} 	&	--0.063	&	0.070	&	0.026	&	7001.90	&	 O~{\sc i} 	&	--0.356	&	0.035	&	0.005	&	 1.282$\mu$m 	&	 H~{\sc i} 	&	--0.767	&	16.183	&	 1.730\\           
5121.83	&	 C~{\sc ii} 	&	--0.065	&	0.080	&	0.053	&	7065.18	&	 He~{\sc i} 	&	--0.364	&	15.063	&	0.677	&	 15.6 $\mu$m 	&	[Ne~{\sc iii}] 	&		&	31.412	&	 1.467\\          
      
\hline
\end{tabular}
\tablenotetext{$^{\ast}$}{stellar origin lines, which have the wide FWHM ($\gtrsim$8 {\AA})}
\end{table*}

\begin{table}
\centering
\caption{Plasma diagnostics.\label{diano_table}}
\begin{tabular}{@{}l@{\hspace{4pt}}r@{\hspace{4pt}}l@{\hspace{5pt}}l@{}}
\hline\hline
Parameter                 &ID&Diagnostic&Result\\
\hline
$n_{\epsilon}$  &(1)&[S\,{\sc ii}] ($\lambda\lambda$6716/31)/($\lambda\lambda$4069/76)&35\,600$\pm$4900\\
(cm$^{-3}$)     &(2)&[O\,{\sc ii}] ($\lambda$3727)/($\lambda\lambda$7320/30)&71\,300$\pm$3900$^{\dagger}$\\
                &(3)&[Ar\,{\sc iv}] ($\lambda$4740)/($\lambda$7170+$\lambda$7263)&139\,500$\pm$60\,400\\
                &(4)&[Cl\,{\sc iii}] ($\lambda$5517)/($\lambda$5537)&50\,700-149\,300\\
\hline
$T_{\epsilon}$ &(5)&[N\,{\sc ii}] ($\lambda\lambda$6548/83)/($\lambda$5755)&10\,600$\pm$1100$^{\ddagger}$\\
(K)            &(6)&[O\,{\sc iii}] ($\lambda$4959+$\lambda$5007)/($\lambda$4363)&11\,900$\pm$200\\    
               &(7)&[Ne\,{\sc iii}] ($\lambda$15.5$\mu$m)/($\lambda$3869)&11\,100$\pm$200\\
               &(8)&[Ar\,{\sc iii}] ($\lambda$7135+$\lambda$7751)/($\lambda$5192)&10\,200$\pm$400\\
\cline{2-4} 
 & &He\,{\sc i} ($\lambda$7281)/($\lambda$6678)&12\,500$\pm$900\\
 & &He\,{\sc i} ($\lambda$7281)/($\lambda$5876)&14\,600$\pm$3300\\
 & &He\,{\sc i} ($\lambda$6678)/($\lambda$4471)&7400-9200\\
 & &He\,{\sc i} ($\lambda$6678)/($\lambda$5876)&11\,200$\pm$1200\\
\cline{2-4} 
 & &(Balmer Jump)/(H 11)&14\,500$\pm$2200\\
\hline
\end{tabular}
\tablenotetext{$^{\dagger}$}{Corrected for recombination contribution to $[$O\,{\sc ii}$]$ $\lambda\lambda$7320/30.}
\tablenotetext{$^{\ddagger}$}{Corrected for recombination contribution to $[$N\,{\sc ii}$]$ $\lambda$5755.}
\end{table}

\begin{table}
\footnotesize
\centering
\caption{Adopting $T_{\epsilon}$ and $n_{\epsilon}$ for 
CELs ionic abundance calculations.\label{temp_ne_cles}}
\begin{tabular}{clcc}
\hline\hline
Zone&
Ions&
$n_{\epsilon}$(cm$^{-3}$)&
$T_{\epsilon}$(K)\\
\hline
1 &N$^{+}$, O$^{+}$, P$^{+}$, S$^{+}$      &$n_{\epsilon}$({\oii})  &$T_{\epsilon}$({\nii})\\
2 &F$^{+}$, S$^{2+}$, Cl$^{2+}$, Ar$^{2+}$,&$n_{\epsilon}$({\cliii})&$T_{\epsilon}$({\ariii})\\
  &Fe$^{2+}$, Kr$^{2+}$&&\\
3 &O$^{2+}$, Ne$^{2+}$, Cl$^{3+}$, Ar$^{3+}$&$n_{\epsilon}$({\ariv})&$T_{\epsilon}$({\oiii})
\\
\hline
\end{tabular}
\end{table}

\subsection{Ionic Abundances}

\subsubsection{CELs Ionic Abundances}
For calculating ionic abundances from the CELs, we adopted a three zone model for 
Hen2-436. Adopted $T_{\epsilon}$ and $n_{\epsilon}$ combinations for each
zone and ion are listed in Table \ref{temp_ne_cles}. $n_{\epsilon}$({\oii}) 
and $T_{\epsilon}$({\nii}) are adopted for ionic abundance derivations 
with I.P. = 0--13.6 eV (zone 1). $n_{\epsilon}$({\cliii}) and $T_{\epsilon}$({\ariii}) are 
for ions with I.P.=13.6--28 eV (zone 2). $n_{\epsilon}$({\ariv}) 
and $T_{\epsilon}$({\oiii}) are for ions with I.P. $>$ 28 eV (zone 3).

The derived ionic abundances are listed in Table \ref{cel_abund}. 
In the last line of the line series of each
ion, we present the adopted ionic abundances in bold face. These values are obtained by the line intensity weighted 
mean when we detected two or more lines. We estimated 14 ionic abundances by solving a $>$5 level atomic
model.

We detected candidates of [F\,{\sc ii}]$\lambda$4789, [P\,{\sc
ii}]$\lambda$7875, and [Kr\,{\sc iii}]$\lambda$6826 as earlier described.  
The estimations of the F$^{+}$, P$^{+}$, Fe$^{2+}$,
and Kr$^{2+}$ abundances are probably done for the first time. For [P~{\sc ii}], we adopted the 
transition probabilities of Mendoza \& Zeippen (1982), the collisional impacts of Tayal (2004), and 
the level energy listed in Atomic Line List v2.05b12\footnote[9]{see http://www.pa.uky.edu/~peter/newpage/}
For [Kr~{\sc iii}], we adopted the 
transition probabilities of Bi{\'e}mont \& Hansen(1986), the collisional impacts of Schoning (1997), and 
the level energy listed in Atomic Line List v2.05b12.

The enhancements of these elements 
would give constraints to parameters (the third dredge-up efficiency, $^{13}$C pocket mass, and the number of 
thermal pulse, etc.) in nucleosynthesis models of low-mass
stars. The detection of these elements is therefore highly interesting.

For the N$^{+}$ and O$^{+}$ estimations, we subtracted the recombination contamination to 
{\nii}$\lambda$5755 and {\oii}$\lambda\lambda$7320/30 from N$^{2+}$ and O$^{2+}$ lines,
respectively. Hence, we obtained the appropriate abundances, i.e., the
amounts derived from their nebular lines. S$^{+}$ abundances from
{\sii} $\lambda$4068/76 lines are lower than those from {\sii} $\lambda$6717/31 
in adopted electron density. S$^{+}$ abundances derived 
from {\sii} $\lambda$4068/76 lines give more reliable 
values than those from {\sii} $\lambda$6717/31. 
To estimate Fe$^{2+}$ abundance, we solved a 33 level model (from $^{5}D_{3}$ to 
b$^{3}P_{2}$) for [Fe {\sc iii}].

\begin{table}
\centering
\caption{CEL ionic abundances.\label{cel_abund}}
\begin{tabular}{@{}lcccccc@{}}
\hline\hline
{X$^{m+}$}&
{$\lambda_{\rm lab}$}&
{Transition}&
{$I(\lambda)$}&
{$\delta I(\lambda)$}&
{X$^{m+}$/H$^{+}$}&
{$\delta$X$^{m+}$/H$^{+}$}\\
{}&
{({\AA}/$\mu$m)}&
{(lower -- upper)}&
{}&
{}&
{}&
{}\\
\hline
N$^{+}$ & 5754.6 &$^{1}D_{2}$--$^{1}S_{0}$& 7.17(--1) & 1.35(--1) & 3.49(--6) & 1.38(--6) \\ 
        & 6548.0 &$^{3}P_{1}$--$^{1}D_{2}$& 3.59(0) & 1.54(--1) & 3.45(--6) & 7.91(--7) \\ 
        & 6583.5 &$^{3}P_{2}$--$^{1}D_{2}$& 1.09(+1) & 5.04(--1) & 3.54(--6) & 8.16(--7) \\ 
        &  &  &  && {\bf 3.52(--6)} &  {\bf 8.37(--7)} \\ 
O$^{+}$ & 3726.0/28.8 &$^{4}S_{3/2}$-$^{2}D_{3/2,5/2}$& 9.25(0) & 4.24(--1) & 3.02(--5) & 1.01(--5) \\ 
        & 7319.0 &$^{4}D_{5/2}$--$^{2}P_{1/2,3/2}$& 6.39(0) & 3.11(--1) & 3.58(--5) & 1.50(--5) \\ 
        & 7330.0 &$^{4}D_{3/2}$--$^{2}P_{1/2,3/2}$& 5.20(0) & 2.73(--1) & 3.54(--5) & 1.48(--5) \\ 
        &  &  &  &&  {\bf 3.32(--5)} &  {\bf 1.28(--5)} \\ 
O$^{2+}$ & 4363.2 &$^{1}D_{2}$--$^{1}S_{0}$& 1.39(+1) & 5.02(--1) & 1.90(--4) & 1.83(--5) \\ 
         & 4931.8 &$^{3}P_{0}$--$^{1}D_{2}$& 1.51(--1) & 8.35(--3) & 2.71(--4) & 2.01(--5) \\ 
  & 4958.9 &$^{3}P_{1}$--$^{2}D_{2}$& 2.65(+2) & 6.37(0) & 1.86(--4) & 1.01(--5) \\ 
  & 5006.8 &$^{3}P_{2}$--$^{2}D_{2}$& 7.92(+2) & 1.61(+1) & 1.93(--4) & 1.02(--5) \\ 
 &  & & &  &  {\bf 1.91(--4)} &  {\bf 1.03(--5)} \\ 
F$^{+}$ & 4789.5 &$^{3}P_{2}$--$^{1}D_{2}$& 1.06(--1) & 1.82(--2) & {\bf 7.19(--8)} & {\bf 1.50(--8)} \\ 
Ne$^{2+}$ & 3868.8&$^{3}P_{2}$-$^{1}D_{2}$ & 5.36(+1) & 1.97(0) & 3.28(--5) & 2.22(--6) \\ 
  & 15.6 &$^{3}P_{2}$--$^{3}P_{1}$& 3.14(+1) & 1.47(0) & 3.98(--5) & 1.92(--6) \\ 
& &  &  &  &  {\bf 3.54(--5)} &  {\bf 2.10(--6)} \\ 
P$^{+}$ & 7875.0 &$^{1}D_{2}$--$^{1}S_{0}$& 4.21(--2) & 7.69(--3) & {\bf 2.31(--8)} & {\bf 7.37(--9)} \\ 
S$^{+}$ & 4068.6 &$^{4}S_{3/2}$--$^{2}P_{3/2}$& 2.07(0) & 2.45(--1) & 2.22(--7) & 6.17(--8) \\ 
  & 4076.4 &$^{4}S_{3/2}$--$^{2}P_{1/2}$& 8.11(--1) & 2.26(--1) & 2.63(--7) & 9.89(--8) \\ 
  & 6716.4 &$^{4}S_{3/2}$--$^{3}D_{5/2}$& 4.88(--1) & 5.35(--2) & 4.02(--7) & 9.16(--8) \\ 
  & 6730.8 &$^{4}S_{3/2}$--$^{3}D_{3/2}$& 1.10(0) & 5.82(--2) & 4.12(--7) & 8.59(--8) \\ 
 &  &  &  &&  {\bf 2.96(--7)} &  {\bf 7.77(--8)} \\ 
S$^{2+}$ & 6312.1 &$^{1}D_{2}$--$^{1}S_{0}$& 1.63(0) & 9.17(--2) & {\bf 4.59(--6)} & {\bf 7.11(--7)} \\ 
Cl$^{2+}$ & 5517.7 &$^{4}S_{3/2}$--$^{2}D_{5/2}$& 5.70(--2) & 6.81(--3) & 5.33(--8) & 8.42(--9) \\ 
 & 5538.0 &$^{4}S_{2}$--$^{3/2}D_{3/2}$& 1.60(--1) & 1.41(--2) & 5.40(--8) & 7.37(--9) \\ 
 &  &  &  &&  {\bf 5.38(--8)} &  {\bf 7.65(--9)} \\ 
Cl$^{3+}$ & 8046.3 &$^{3}P_{2}$--$^{1}D_{2}$& 1.65(--1) & 1.16(--2) & {\bf 8.11(--9)} & {\bf 6.30(--10)} \\ 
Ar$^{2+}$ & 5191.8 &$^{1}D_{2}$--$^{1}S_{0}$& 6.34(--2) & 5.46(--3) & 6.78(--7) & 1.30(--7) \\ 
  & 7135.8 &$^{3}P_{2}$--$^{1}D_{2}$& 7.66(0) & 3.85(--1) & 6.71(--7) & 6.52(--8) \\ 
  & 7751.1 &$^{3}P_{1}$--$^{1}D_{2}$& 1.84(0) & 9.90(--2) & 6.72(--7) & 6.66(--8) \\ 
 &  &  &  &&  {\bf 6.71(--7)} &  {\bf 6.59(--8)} \\ 
Ar$^{3+}$ & 4740.2&$^{4}S_{3/2}$--$^{2}D_{3/2}$ & 7.25(--1) & 1.89(--2) & 8.58(--8) & 4.22(--9) \\ 
  & 7170.5 &$^{2}D_{3/2}$--$^{2}P_{3/2}$& 3.47(--2) & 6.48(--3) & 8.71(--8) & 1.72(--8) \\ 
  & 7262.7 &$^{2}D_{3/2}$--$^{2}P_{1/2}$& 4.29(--2) & 2.10(--2) & 1.24(--7) & 6.10(--8) \\ 
 &  &  &  & & {\bf 8.79(--8)} &  {\bf 7.81(--9)} \\ 
Fe$^{2+}$&5270.4&$^{5}D_{3}$-$^{3}P2_{2}$&	7.88(--2)&	1.08(--2)&	{\bf 7.91(--8)}&
 {\bf 1.36(--8)}\\
Kr$^{2+}$&6826.4&$^{3}P_{2}$--$^{1}D_{2}$	&3.57(--2)	&4.83(--3)	&{\bf 3.11(--9)}
 &{\bf 5.00(--10)}\\
\hline
\end{tabular}
\tablecomments{We corrected for recombination contributions to
 $[$N\,{\sc ii}$]$ $\lambda$5755 and $[$O\,{\sc ii}$]$ $\lambda\lambda$7320/30.}

\end{table}

\subsubsection{ORLs Ionic Abundances \label{ir}}
The estimated ionic abundances derived from ORLs are listed in Table \ref{rec_hec}. 
This work may constitute the first ever estimation of the N and O ORL abundance. The ORL ionic abundances are derived
from 
\medskip

\begin{equation}
\label{abunr}
\frac{N({\rm X^{m+}})}{N({\rm H^{+}})} = \frac{\alpha({\rm
H\beta})}{\alpha({\rm X^{m+}})}
\frac{\lambda({\rm X^{m+}})}{\lambda({\rm H\beta})}\frac{I({\rm
X^{m+}})}{I({\rm H\beta})},
\end{equation}
\medskip

\noindent
where $\alpha({\rm X^{m+}})$ is the recombination coefficient for the
ion ${\rm X^{m+}}$. From the log$n_{\epsilon}$--$T_{\epsilon}$ 
plot of the {\oii} $\lambda$3727/$\lambda\lambda$7320/30 ratio and 
estimated values of $T_{\epsilon}$(BJ) and $T_{\epsilon}$({\hei}), we calculate the electron density to be $\sim$10$^{5}$ cm$^{-3}$. 
We adopted this value for all ORL ionic abundance calculations. We adopted $T_{\epsilon}$(BJ) 
for all recombination lines except for He$^{+}$. We adopted $T_{\epsilon}$(He\,{\sc i}) for He$^{+}$ abundances.

Effective recombination coefficients for the lines' parent multiplets
were taken from the references listed in Table 11 of Otsuka et al. (2010). 
The recombination coefficient of each line was obtained by a branching 
ratio, $B(\lambda_{i})$, which is the ratio of the recombination 
coefficient of the target line, 
$\alpha(\lambda_{i})$ to the total recombination coefficient, 
$\sum_{i} \alpha(\lambda_{i})$ in a multiplet line. 
To calculate the branching ratio, we referred to Wiese et al. (1996)
except for the O\,{\sc ii} $3d$-$4f$ transition line O\,{\sc ii} $\lambda$4292.2. For this line, 
the branching ratios were provided by Liu et al. (1995) based on an 
intermediate coupling. We applied the Case B assumption (Baker \& Menzel 1938) for lines of levels 
with the same spin as the ground state 
and the Case A (optically thin in Ly-$\alpha$) assumption for the other multiplicity lines. In the last line of the line series 
of each ion, we present the adopted ionic abundance and the error 
estimated by the weighted mean of the line intensity. In the following, we give 
short comments to several ionic abundance estimations.

The He$^{+}$ abundances were estimated using 5 different 
{\hei} lines selected to be insensitive to electron density in order to reduce intensity enhancement by collisional 
excitation from the He$^{0}$ 2$s$ $^{3}S$ level. The collisional excitation 
from the He$^{0}$ 2$s$ $^{3}S$ level mainly enhances the intensity of the triplet {\hei}
lines. We removed this contribution using the formulae given by Kingdon \& Ferland (1995). The collisional
excitation contamination for our measured lines was estimated to be 14.8 $\%$ for 6678 {\AA},
20.5 $\%$ for 4471 {\AA}, 7.5 $\%$ for 4388 {\AA}, 9.7 $\%$ for 4921
{\AA}, and 37.9 $\%$ for 5876 {\AA}. The derived He$^{+}$
abundances well agreed with each other within error.

We detected several C\,{\sc iii} lines. Of them, we chose C\,{\sc iii}
$\lambda$8196 to estimate the C$^{3+}$ abundance. Its line width is narrow
(3.97 {\AA}), therefore this line could be of nebular origin. Since the ground state
of C\,{\sc iii} is singlet (2$s^{2}$ $^{1}S$), we adopted the Case A
assumption for this line. We detected several C\,{\sc iv} 
and O\,{\sc iii} lines of wide line width. Since they would be
of stellar wind origin, we did not estimate their ionic abundances; Hen2-436 could not be a
highly ionized nebula, because the high I.P. nebular lines such as 
{\neiv} were unseen in the FORS2 spectra. The central star of the PN is 
not hot enough to ionize these lines.   

We measured O\,{\sc ii} $\lambda$4292.2 (doublet, 3$d$-4$f$) and
$\lambda$4676.2 (quadruplet, 3$s$-3$p$) lines as well. {O\,{\sc ii} $\lambda$4291.2 
is blended with {O\,{\sc ii} $\lambda\lambda$4291.86,4292.98 {\AA}.
Since the ground level of O\,{\sc ii} 
is quadruplet, we adopted Case A for the doublet and Case B 
for the quadruplet lines. We found that the discrepancy between O$^{2+}$ ORL 
and CEL abundances is $>$1 dex. The O$^{2+}$ discrepancy has been found in many PNe including 
BoBn1 (see section 4 of Otsuka et al. 2010).

\begin{table}
\centering
\caption{ORL ionic abundances.\label{rec_hec}}
\begin{tabular}{@{}lccccc@{}}
\hline\hline
{X$^{m+}$}&
{$\lambda_{\rm lab}$}&
{$I(\lambda)$}&
{$\delta$$I(\lambda)$}&
{X$^{m+}$/H$^{+}$}&
{$\delta$X$^{m+}$/H$^{+}$}\\
{}&
{({\AA})}&
{}&
{}&
{}&
{}\\
\hline
He$^{+}$ & 6678.2 & 4.29(0) & 1.69(--1) & 1.06(--1) & 1.06(--2) \\ 
 & 4471.5 & 4.91(0) & 1.24(--1) & 1.33(--1) & 1.36(--2) \\ 
 & 4387.9 & 6.24(--1) & 3.58(--2) & 1.06(--1) & 1.31(--2) \\ 
 & 4921.9 & 1.50(0) & 3.71(--2) & 1.09(--1) & 9.93(--3) \\ 
 & 5875.6 & 1.42(+1) & 3.96(--1) & 1.12(--1) & 1.29(--2) \\ 
 &  &  &  & {\bf 1.15(--1)} & {\bf 1.25(--2)} \\ 
C$^{2+}$ & 4267.2 & 1.10(0) & 3.16(--2) & 1.14(--3) & 2.65(--4) \\ 
 & 6151.3 & 5.05(--2) & 8.44(--3) & 1.14(--3) & 2.99(--4) \\ 
 & 7231.3 & 3.45(--1) & 3.86(--2) & 8.04(--4) & 2.47(--4) \\ 
 & 7236.4 & 5.59(--1) & 3.56(--2) & 6.52(--4) & 1.91(--4) \\ 
 &  &  &  & {\bf 9.50(--4)} & {\bf 2.43(--4)} \\ 
C$^{3+}$ & 8196.5 & 4.57(--2) & 5.60(--3) & {\bf 6.97(--5)} & {\bf 1.72(--5)} \\ 
N$^{2+}$ & 5480.1 & 4.56(--2) & 3.24(--3) & 5.69(--4) & 1.22(--4) \\ 
 & 5927.8 & 1.57(--2) & 3.81(--3) & 5.30(--4) & 2.20(--4) \\ 
 &  &  &  & {\bf 5.59(--4)} & {\bf 1.47(--4)} \\ 
O$^{2+}$ & 4292.2 & 3.98(--2) & 2.36(--2) & 1.61(--3) & 1.01(--3) \\ 
 & 4676.2& 3.39(--1) & 2.94(--2) & 3.30(--3) & 6.81(--4) \\ 
 &  &  &  & {\bf 3.12(--3)} & {\bf 7.15(--4)}  \\
\hline
\end{tabular}
\tablecomments{For He$^{+}$, the contribution of collisional excitation
 was subtracted. O\,{\sc ii}
	    $\lambda$4291.26,4291.86,4292.21,4292.98 are observed as a blended line (See the text).}
\end{table}

\subsection{Elemental Abundances}
To estimate elemental abundances, we estimated the 
unobserved ionic abundances 
using an ionization correction factor, ICF(X). ICFs(X) for
each element are listed in Table \ref{icf}.

The He abundance is the sum of He$^{+}$ and He$^{0+}$ abundances, and we estimated the  
unseen He$^{0}$ abundance.  The C abundance is the sum of the C$^{+}$, C$^{2+}$, C$^{3+}$ abundances. 
We corrected for the unseen C$^{+}$ assuming (C$^{+}$/C) = (N$^{+}$/N)$_{\rm
CELs}$. The N abundance is the sum of N$^{+}$ and N$^{2+}$. For the CEL N abundance, 
N$^{2+}$ was unobserved and calculated using Table~\ref{icf}. For the ORL N abundance, we accounted for the unseen N$^{+}$ assuming (N/N$^{+}$)$_{\rm ORLs}$ = 
(Ar/Ar$^{2+}$)$_{\rm CELs}$. The O abundance is the sum of the O$^{+}$ and O$^{2+}$ abundances. 
For the ORL O abundance, we assumed (O$^{2+}$/O)$_{\rm ORLs}$ = (O$^{2+}$/O)$_{\rm CELs}$. 
The Ne abundance is the sum of the Ne$^{+}$ and Ne$^{2+}$ abundances, and we corrected for the unseen Ne$^{+}$. 
Assuming that the F abundance is the sum of the F$^{+}$ and F$^{2+}$ abundances, we estimated the unobserved F$^{2+}$ abundance. 
The S abundance is the sum of the S$^{+}$, S$^{2+}$, and S$^{3+}$ abundances. We accounted for the unseen 
S$^{3+}$ abundance using the CEL O and O$^{+}$ abundances. We assume that the Cl abundance is the sum of Cl$^{+}$, Cl$^{2+}$, 
and Cl$^{3+}$ abundances. The unseen Cl$^{+}$ is accounted for by assuming (Cl/Cl$^{+}$) = (O/O$^{+}$)$_{\rm CELs}$. 
The Ar abundance is the sum of the Ar$^{2+}$ and Ar$^{3+}$ abundances, and we accounted for the unseen Ar$^{+}$. 
The Fe abundance is the sum of the Fe$^{2+}$ and Fe$^{3+}$ abundances, and we accounted for the unseen Fe$^{3+}$.
The Kr abundance is the sum of the Kr$^{+}$, Kr$^{2+}$ and Kr$^{3+}$ abundances, and we accounted for the unseen Kr$^{+}$ and Kr$^{3+}$.

The resultant elemental abundances are listed in Table \ref{element}. In the third column, 
the number densities relative to H are listed. In the fourth column, the logarithmic 
number densities are given when $\log_{10}$(H) is 12. In the fifth, sixth, and seventh columns, we present the 
enhancements relative to the solar abundances. For the solar abundances, we adopted the 
amounts recommended by Lodders (2003). In the last two columns, we present the number density 
in the form of $\log_{10}$(X/H)+12 estimated by Walsh et al. (1997) and Dudziak et al. (2000). 
The amounts of Walsh et al. (1997) were estimated based on optical spectra and those of 
Dudziak et al. (2000) by a
photo-ionization model. Our estimated amounts agree fairly well with their values 
except N and S. The overabundances of 
these elements relative to Walsh et al. (1997) would be due to the adopted $T_{\epsilon}$ 
for N$^{+}$, S$^{+}$ and S$^{2+}$estimations. Walsh et al. (1997) adopted 
12\,600 K for all ionic abundance estimations, while we adopted 10\,600 K 
for N$^{+}$ and S$^{+}$. If we adopted a $T_{\epsilon}$ of 12\,600 K, our N and S abundances would decrease 
by --0.17 and --0.54 dex, respectively.

Our estimated [Cl,Ar/H] are almost consistent within the estimated error. Since both elements are 
not synthesized in low-mass stars and do not couple with the dust grains, they would indicate 
the metallicity of the progenitor. The metallicity of Hen2-436 ($\sim$--0.5) is comparable 
to a typical LMC metallicity. The large depletion of the Fe abundance ([Fe/H]) indicates that most of 
the Fe exists not as ionized gas but as dust grains.

The discrepancy between the ORL and CEL O abundances ($>$1 dex) is not due to the 
adopted electron temperatures; the adopted $T_{\epsilon}$ for the ORL O$^{2+}$ estimations is 
consistent with that for CEL O$^{2+}$. The O ORLs might be emitted
from O-rich and H-deficient knots, similar to those observed in Abell 30 (e.g., Wesson et al. 2003) because the central star of Hen2-436
is a [WC] star and most stars of this type are H-deficient. According
to the classification by Acker \& Neiner (2003), the central star of Hen2-436 is [WC4-6] based on 
the instrumental broadening subtracted FWHM of C\,{\sc iv} $\lambda$5812 (29.6 {\AA}) and the line 
ratios of C\,{\sc ii,iii} to C\,{\sc iv} $\lambda$5812. Girard et al. (2007) classified Hen2-436 
as [WC4]. In this paper, we will not examine the O discrepancy problem further. 
To resolve the O abundance discrepancy, it is necessary to obtain high-dispersion spectra 
to increase the chance of O\,{\sc ii} detections and properly de-blend these O~{\sc ii} lines with the others. 

We are able to estimate the C/O ratio of Hen2-436 to be 0.54 $\pm$ 0.41 from the elemental ORL C and O abundances. 
This is the first time the C/O ratio has been calculated for this nebula
using the same type of emission line.  
In Hen2-436, the CEL C abundance has not previously been estimated as
mentioned earlier. So far, both the CEL and ORL C abundances in PNe have been estimated 
by Wang \& Liu (2007), Wesson et al. (2005), Liu et al. (2004), and Tsamis et al. (2004). From these works, 
we found the C$^{2+}$(ORL)/C$^{2+}$(CEL) of 4.10 $\pm$ 0.49 among 58 PNe and the C$^{3+}$(OLR)/C$^{3+}$(CEL) of 4.52 $\pm$ 0.85 among 14 PNe. Using 
those factors, the CEL C$^{2+}$ and C$^{3+}$ abundances are extrapolated to be 2.32(--4) $\pm$ 6.34(--5) and 1.54(--5) $\pm$ 4.78(--6), respectively. 
The expected CEL C abundance is 2.89(--4) $\pm$ 1.95(--4) and the CEL C/O is 1.68 $\pm$ 1.08 when we adopt the same ICF listed in Table \ref{icf}.

The F, P, and Kr abundances were estimated. These
elements are known to be synthesized by neutron capture processes in the He-rich intershell region
during the thermally pulsing AGB phase (e.g., Abia et al. 2010 
and Lugaro et al. 2004 for F production, Herwig 2005 and Busso et al. 1999 for AGB nucleosynthesis including the {\it s}-process). Concerning low-mass 
stars ($<$4 $M_{\odot}$), most of the free neutrons are released by the 
$^{13}$C($\alpha$,$n$)$^{16}$O reaction 
in the interpulse phase. $^{13}$C isotope is produced by partial mixing of the bottom of the 
H-rich convective envelope into the outermost region of the $^{12}$C-rich intershell layer. 
The nucleosynthesis models for 
low- to intermediate mass stars predict that C and neutron capture elements produced 
in the He-rich intershell are brought up to the stellar surface by the third dredge up. 
Figure \ref{cf} (a), (b), and (c) shows the relations between [C/Ar] and [F/Ar], [P/Ar], and [Kr/Ar] 
among PNe and C-rich stars, respectively. The trends between [C/Ar] and [F,P,Kr/Ar] 
support the theoretical description of the {\it s}-process and the third dredge-up occurring in AGB stars.

Among Sgr dwarf galaxy PNe, F lines have been detected in the halo PN BoBn1 (Otsuka et al. 2010, 2008), which is 
also suspected to be an old Sgr dwarf galaxy PN. Hen2-436 would be the second detection case.

The only stable P isotope is $^{31}$P. $^{31}$P is synthesized in the He-rich intershell via the following reaction.

\begin{eqnarray}
^{26}{\rm Mg}(n,\gamma)^{27}{\rm Mg}(\beta^{+})^{27}{\rm Al}(n,\gamma)^{28}{\rm Al}(\beta^{+}) \nonumber\\
^{28}{\rm Si}(n,\gamma)^{29}{\rm Si}(n,\gamma)^{30}{\rm Si}(n,\gamma)^{31}{\rm Si}(\beta^{+})^{31}{\rm P} 
\end{eqnarray}

In Table \ref{pabund}, we list the P abundances estimated for 9 PNe. 
The listed P abundances are of nebular origin and they are estimated using [P~\,{\sc ii}]$\lambda\lambda$4669,7875 
(NGC6572, NGC6741, NGC7027, and Hen2-436), [P\,{\sc ii}]$\lambda\lambda$1.14/1.19 $\mu$m (IC5117 and 
NGC7027), [P\,{\sc iii}]$\lambda$17.89 $\mu$m (NGC40, NGC2392, NGC6210, and NGC6826). We did not detect 
[P\,{\sc ii}]$\lambda\lambda$1.14/1.19 $\mu$m in the MMIRS spectrum due to the low S/N.

The 2.5 $M_{\odot}$ models of Karakas et al. (2009) predict that the P abundance 
is not largely enhanced after AGB nucleosynthesis ($\lesssim$0.1 dex) with or without a partial mixing zone. The P production 
in the He-rich intershell depends on the amount of extra mixing included in the calculations to produce the $^{13}$C pocket 
(Werner \& Herwig 2006). According to Werner \& Herwig (2006), P has been detected in a number of PG 1159 stars by the identification of 
the P~{\sc v} $\lambda\lambda$1118/28 lines, and their P abundances are roughly solar. PG 1159 stars are H-deficient.
 Marcolino et al. (2007) observed 4 PNe with WR type central stars using the far-UV spectrograph $FUSE$, and they found P~{\sc v} 
lines in BD+30$^{\circ}$3639, NGC40, and NGC5315. They argued that the P abundance in NGC5315 is 4-25
times the solar abundance and the value agrees with nucleosynthesis calculations of Werner \& Herwig (2006). For the other PNe, 
they did not give the P abundances. Our estimated [P/H] abundance could agree with  Karakas et al. (2009) and Werner \& Herwig (2006). 
However, except for Hen2-436 and NGC40, PNe listed in Table \ref{pabund} do not have WR central stars, and 4 PNe of these no-WR 7 
PNe show $>$0.2 dex P enhancement. To verify the discrepancy of P enhancement between the observation and the models, we 
need to increase the detection cases, for WR-type PNe in particular. We should note that the production of P in AGB models 
is strongly dependent on a number of modeling uncertainties including reaction rates, the adopted convective mixing model, 
and the mass-loss rate; for example, Karakas \& Lattanzio (2007) and Karakas (2010) adopted different reaction 
rates  from each other, and the models by the former produced about a factor of 10 more P than those of the latter. 
The P abundance could give a constraint to uncertain parameters in the AGB nucleosynthesis models, including 
the $^{13}$C pocket mass, and it would be an efficient diagnostic tool, because P lines can be observed from the far-UV to the mid-IR.

Among PNe in which F, P, and Kr have been detected, Hen2-436 seems to be 
very similar to NGC40 in elemental abundances, dust composition, and central star properties. The elemental abundances of NGC40 are listed in
the last column of Table \ref{element}. In terms of elemental abundances, the C/O ratio and the [Cl,Ar,Fe/H] abundances 
are comparable to those of Hen2-436 as present in Table \ref{element}. The ORL and CEL C/O ratios of NGC40 are 0.46 and 1.41, respectively. 
The C(ORL)/O(CEL) ratio is 8.32 (7.25 $\pm$ 4.86 in Hen2-436).
F, P, and Kr abundances are also comparable. In addition, the enhancements of ORL N
and O abundances are relatively similar to those of CEL abundances. 
Liu et al. (2004) reported that the C$^{2+}$(ORL)/C$^{2+}$(CEL) and O$^{2+}$(ORL)/O$^{2+}$(CEL) ratios are 
5.8 and 17.8 (16.3 $\pm$ 3.9 in Hen2-436), respectively. 
Ramos-Larios et al. (2010) found PAHs emissions in the {\it ISO}/SWS and {\it Spitzer}/IRS spectra of NGC40. Since our estimated CEL and ORL 
C/O ratios of Hen2-436 are larger than the PAH 7.7 $\mu$m band detection limit by Cohen \& Barlow (2005) 
and the chemistry of Hen2-436 seems to be similar to NGC40, Hen2-436 might have PAHs.
The central stars of both PNe have wide FWHM C\,{\sc iii}, C\,{\sc iv}, and N\,{\sc iii} stellar origin
lines, and they have WR-type central stars. De Marco \& Barlow (2001) classified NGC40 as [WC8]. The strong wind could contribute to 
enhancements of C, F, P, and Kr abundances in the nebula.

\begin{figure*}
\begin{tabular}{c@{\hspace{-0.5pt}}c@{\hspace{-0.5pt}}c}
\includegraphics[scale=0.375,clip]{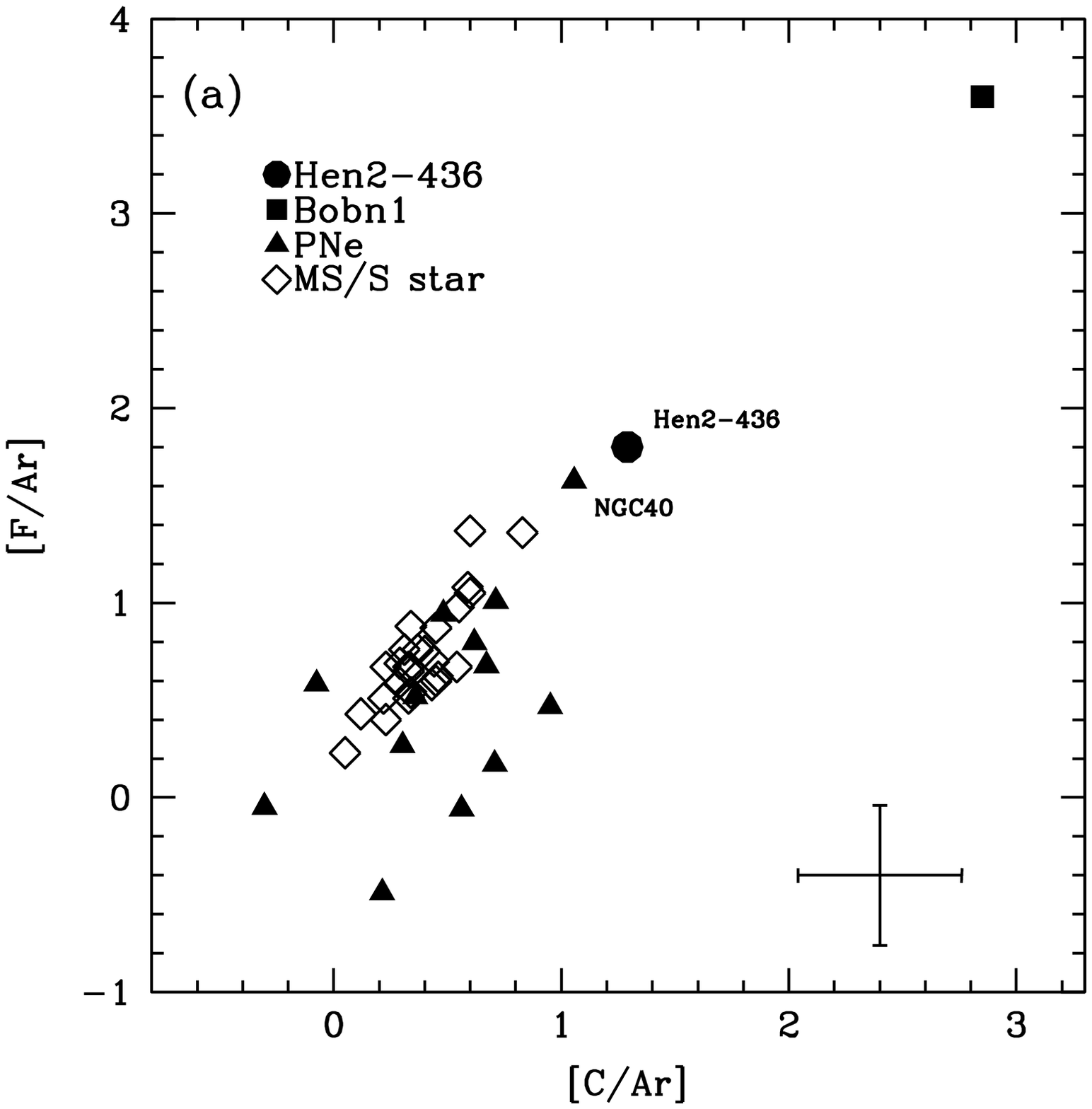}&
\includegraphics[scale=0.375,clip]{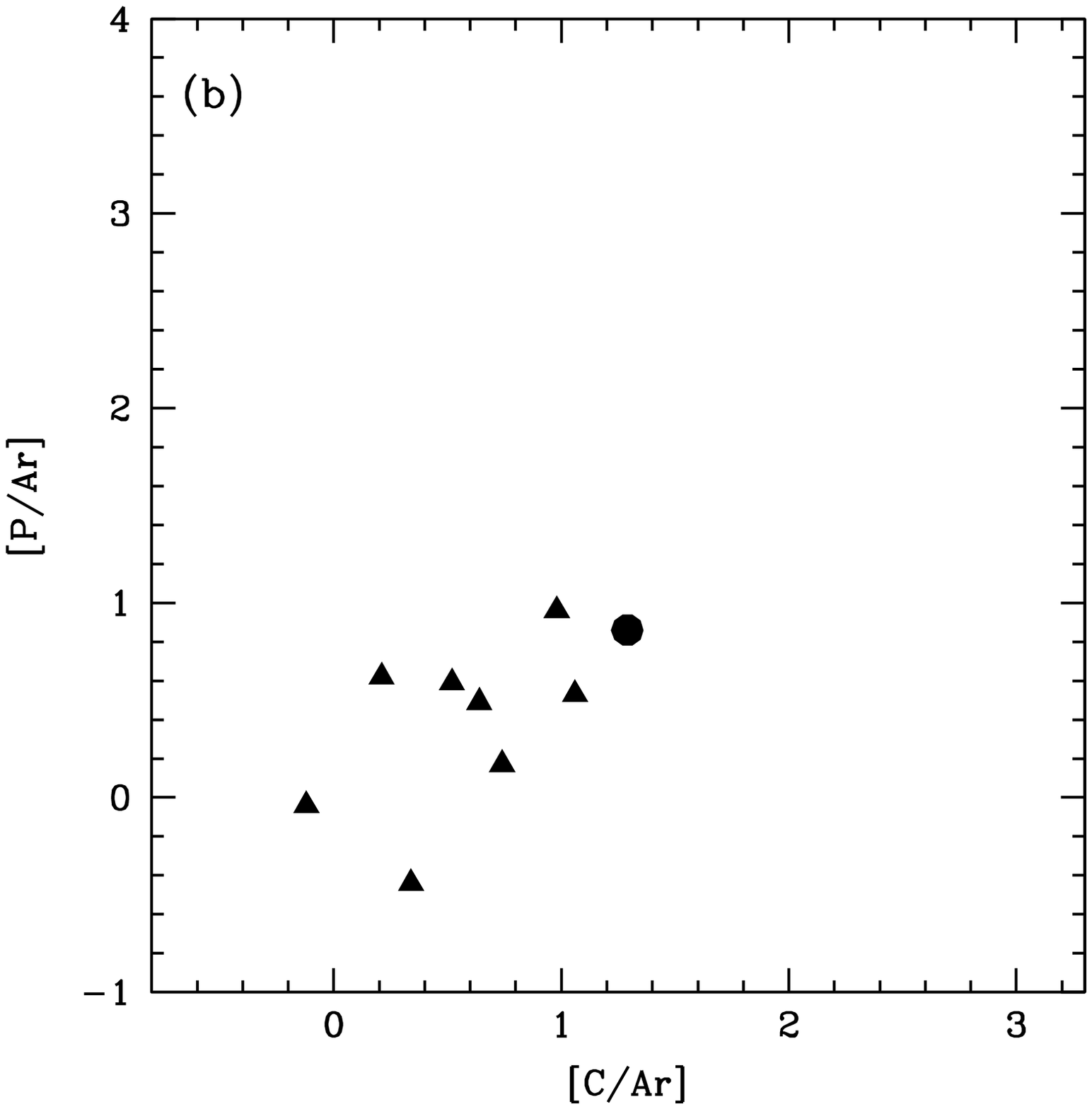}&
\includegraphics[scale=0.375,clip]{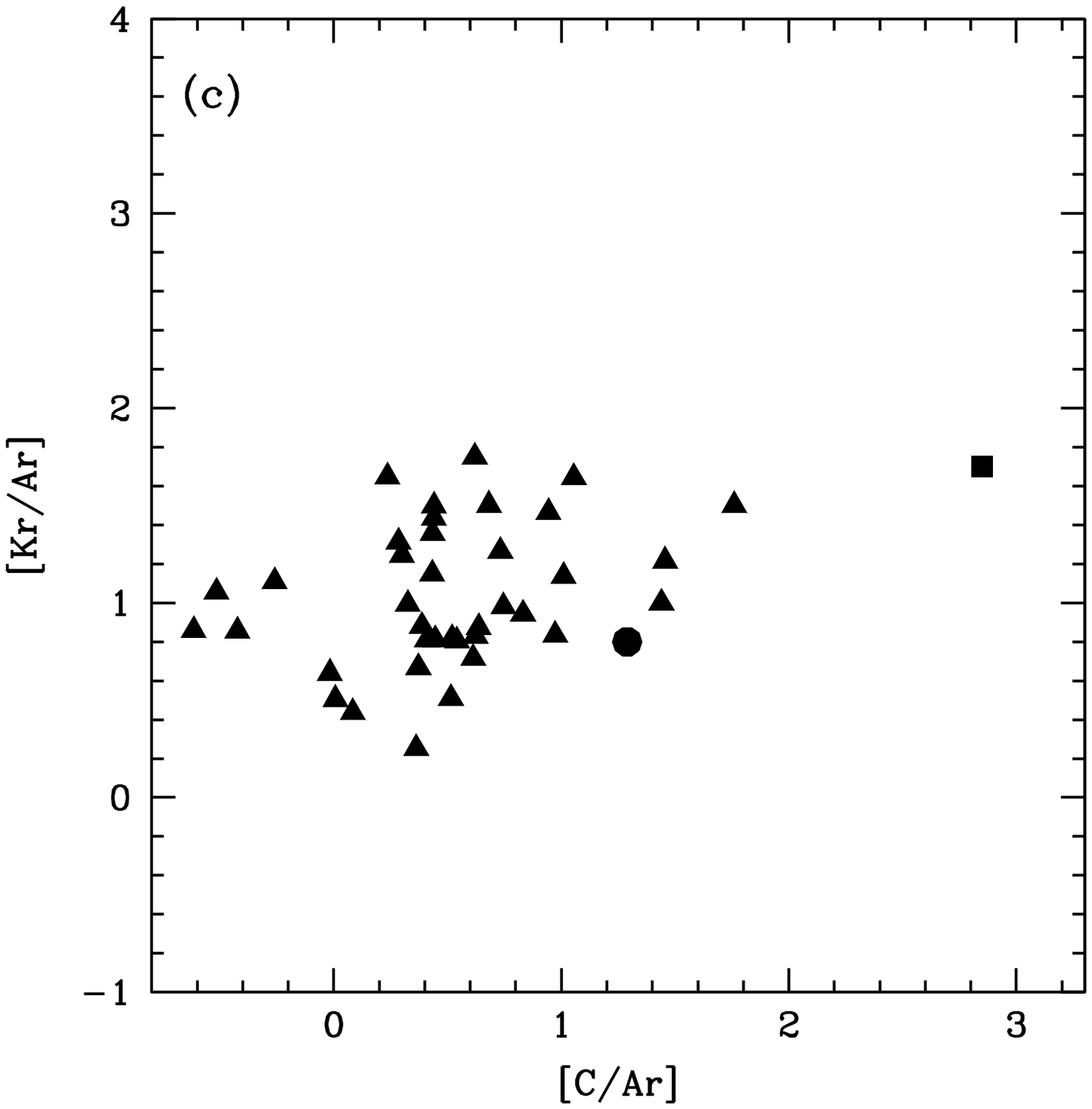}
\end{tabular}
\caption{({\it left panel}): The [F/Ar]-[C/Ar] diagram. The typical error is indicated by the cross (the same for the other diagrams)
. The data of BoBn1 are from Otsuka et al. (2010). 
The references for Galactic PNe and MS/S stars are in Otsuka et al. (2008). ({\it middle panel}): 
The [P/Ar]-[C/Ar] diagram. The references for Galactic PNe are in Table \ref{pabund}. 
({\it right panel}): The [Kr/Ar]-[C/Ar] diagram. The data for BoBn1 (upper limit)
is from Otsuka et al. (2010). The data for Galactic PNe are taken from Sterling \& Dinerstein (2008).
\label{cf}}
\end{figure*}

\begin{table}
\centering
\caption{Adopted Ionization Correction Factors (ICFs).\label{icf}}
\begin{tabular}{@{}llll@{}}
\hline\hline
{X}&
{Line}&
{ICF(X)}&
{X/H}\\
\hline
He &ORLs &(S$^{+}$+S$^{2+}$)/S$^{2+}$&ICF(He)He$^{+}$\\
\noalign{\medskip}

C  &ORLs &(1-(N$^{+}$/N))$^{-1}$&ICF(C)(C$^{2+}$+C$^{3+}$)\\ 
\noalign{\medskip}

N &CELs &(O/O$^{+}$) &ICF(N)N$^{+}$\\
  &ORLs &(Ar/Ar$^{2+}$)&ICF(N)N$^{2+}$\\
\noalign{\medskip}

O  &CELs & 1 &O$^{+}$+O$^{2+}$\\
   &ORLs &(O/O$^{2+}$)$_{\rm CELs}$ &ICF(O)O$^{2+}$\\
\noalign{\medskip}

F &CELs &(O/O$^{+}$)&ICF(F)F$^{+}$\\
\noalign{\medskip}

Ne &CELs &(O/O$^{2+}$)$_{\rm CELs}$ &ICF(Ne)Ne$^{2+}$\\
\noalign{\medskip}

P &CELs &(S/S$^{+}$)&ICF(P)P$^{+}$\\
\noalign{\medskip}

S &CELs &$\left[1 - (1-({\rm O^{+}/O}))^{3}\right]^{-1/3}$ &ICF(S)(S$^{+}$+S$^{2+}$)\\
\noalign{\medskip}

Cl &CELs &(1-(O$^{+}$/O))$^{-1}$ &ICF(Cl)$\rm \left(Cl^{2+}+Cl^{3+}\right)$\\
\noalign{\medskip}

Ar &CELs &(1-(N$^{+}$/N))$^{-1}$&ICF(Ar)(Ar$^{2+}$+Ar$^{3+}$)\\
\noalign{\medskip}

Fe &CELs &(O/O$^{+}$)&ICF(Fe)Fe$^{2+}$\\
\noalign{\medskip}

Kr &CELs &Cl/Cl$^{2+}$ &ICF(Kr)Kr$^{2+}$\\
\hline
\end{tabular}
\end{table}

\begin{table*}
\centering
\footnotesize
\caption{Elemental abundances of Hen2-436.\label{element}}
\begin{tabular}{@{}cccrcccrrr@{}}
\hline\hline
{X}&
{Line}&
{X/H}&
{log(X/H)+12}&
{$[$X/H$]$}&
{$[$X/Ar$]$}&
{$[$X/O$]^{\dagger}$}&
{Ref.(1)}&
{Ref.(2)}&
{Ref.(3)}\\
{}&
{}&
{}&
{(H=12)}&
{}&
{}&
{}&
{Hen2-436}&
{Hen2-436}&
{NGC40}\\
\hline
He&ORLs& 1.22(--1)$\pm$1.35(--2)&11.09$\pm$0.05&+0.19$\pm$0.05&+0.79$\pm$0.33&+0.53$\pm$0.08&	11.02$\pm$0.02&11.03$\pm$0.01 &11.08\\
C&CELs&$\cdots$&$\cdots$&$\cdots$&$\cdots$&$\cdots$&$\cdots$&$\cdots$&8.84\\
&ORLs&1.20(--3)$\pm$7.97(--4)&	9.08$\pm$0.35&+0.69$\pm$0.35&+1.29$\pm$0.48&+1.03$\pm$0.36&$\cdots$&9.06$\pm$0.09&9.61\\
N&CELs&2.37(--5)$\pm$1.09(--5)&	7.38$\pm$0.22&--0.45$\pm$0.24&+0.15$\pm$0.41&--0.11$\pm$0.25&6.97$\pm$0.16&7.42$\pm$0.06&7.93\\
&ORLs&7.42(--4)$\pm$5.10(--4)&	8.87$\pm$0.37&	+1.04$\pm$0.38&	+1.64$\pm$0.51&	+1.38$\pm$0.39&	$\cdots$&	$\cdots$&9.32	\\
O&CELs&2.24(--4)$\pm$1.64(--5)&	8.35$\pm$0.03&--0.34$\pm$0.06&+0.26$\pm$0.34&+0.00$\pm$0.08&8.29$\pm$0.08&8.36$\pm$0.06&8.69\\
&ORLs&3.67(--3)$\pm$9.04(--4)&	9.56$\pm$0.11&	+0.87$\pm$0.12&	+1.47$\pm$0.35&	+1.21$\pm$0.13&	$\cdots$&	$\cdots$&9.95	\\
F&CELs&4.85(--7)$\pm$2.15(--7)&	5.69$\pm$0.21&	+1.23$\pm$0.22&	+1.83$\pm$0.39&	+1.57$\pm$0.22&	$\cdots$&	$\cdots$&5.48	\\
Ne&CELs&4.15(--5)$\pm$4.51(--6)&	7.62$\pm$0.05&	--0.25$\pm$0.11&	+0.35$\pm$0.35&	+0.09$\pm$0.13&7.57$\pm$0.08&	7.54$\pm$0.06&8.01\\
P&CELs&5.26(--7)$\pm$2.38(--7)&	5.72$\pm$0.21&	+0.26$\pm$0.22&	+0.86$\pm$0.39&	+0.60$\pm$0.22&	$\cdots$&	$\cdots$&5.38	\\
S&CELs&6.73(--6)$\pm$1.23(--6)&	6.83$\pm$0.08&	--0.36$\pm$0.09&	+0.24$\pm$0.34&	--0.02$\pm$0.11&6.30$\pm$0.08&	6.59$\pm$0.05&6.41\\
Cl&CELs&7.26(--8)$\pm$1.48(--8)&	4.86$\pm$0.09&	--0.47$\pm$0.11&	+0.13$\pm$0.35&	--0.13$\pm$0.12&$\cdots$&$\cdots$&4.91	\\
Ar&CELs&8.91(--7)$\pm$5.59(--7)&	5.95$\pm$0.32&	--0.60$\pm$0.33&	+0.00$\pm$0.47&	--0.26$\pm$0.34&5.76$\pm$0.12&5.78$\pm$0.08&5.94\\
Fe&CELs&5.34(--7)$\pm$2.28(--7)&	5.73$\pm$0.20&	--1.74$\pm$0.20&	--1.14$\pm$0.39&	--1.40$\pm$0.21&$\cdots$&$\cdots$&5.79	\\
Kr&CELs&4.21(--9)$\pm$1.24(--9)&3.62$\pm$0.13&+0.26$\pm$0.15&+0.86$\pm$0.36&+0.60$\pm$0.17&$\cdots$&$\cdots$&$>$4.19	\\	
\hline
\end{tabular}
\tablenotetext{$^{\dagger}$}{For Hen2-436, we adopted the CEL O abundance.}
\tablerefs{(1) Walsh et al. (1997) and (2) Dudziak et al. (2000) for elemental abundances of Hen2-436. (3) Pottasch et al. (2003) for P and Fe, Zhang \& Liu (2005) for F, Sterling \& Dinerstein (2008) for Kr, and Liu et al. (2004) for the others of NGC40.}
\end{table*}

\begin{table}
\centering
\caption{P Abundances in 9 PNe}
\begin{tabular}{lccccl}
\hline\hline
Nebula	&[P/H]	&[Ar/H]	&[P/Ar]	&[C/Ar]&Ref.\\
\hline
IC5117  &+0.14  &--0.34 &+0.49 &+0.64 &(1),(2)\\
NGC40	&--0.08	&--0.61	&+0.53	&+1.06$^{\dagger}$&(3)\\
NGC2392	&--0.65	&--0.21	&--0.44	&+0.34&(4)\\
NGC6210	&--0.23	&--0.19	&--0.04	&--0.12&(5)\\
NGC6826	&--0.23	&--0.40	&+0.17	&+0.74&(6)\\
NGC6572	&+0.34	&--0.27	&+0.62	&+0.21&(7)\\
NGC6741	&+0.58	&--0.01	&+0.59	&+0.52&(8)\\
NGC7027	&+0.71	&--0.25	&+0.96	&+0.98&(9)\\
Hen2-436&+0.26	&--0.60	&+0.86	&+1.29&(10)\\
\hline
\end{tabular}
\tablenotetext{$^{\dagger}$}{We adopted the CEL C abundance.}
\tablerefs{(1) Rudy et al. (1991) for the P abundance; (2) Hyung et al. (2001) for the other elements; 
(3) see references in Table \ref{element}.; (4) Pottasch et al. (2008); (5) Pottasch et al. (2009);
 (6) Surendiranath \& Pottasch (2008); (7) Hyung et al. (1994); (8) Hyung \& Aller (1997); (9) Otsuka in prep.; (10) This work. \label{pabund}}
\end{table}

\section{Discussion}

\subsection{Photo-Ionization Modeling}
To investigate the properties of the ionized gas, dust, 
and the central star of the PN in a self-consistent 
way, we constructed a theoretical photoionization (P-I) model which matches 
the observed flux of emission lines and the spectral energy 
distribution (SED) between optical and mid-IR wavelengths, using 
{\sc Cloudy} c08.00 (Ferland 2004).

To construct an accurate P-I model, we require information about the distance from us 
to the object, the incident SED from the central star and the 
elemental abundances, geometry, density distribution, and size 
of the nebula. Since Hen2-436 is in the core of the Sagittarius dwarf galaxy as we 
mentioned in the introduction, we fix the distance at 24.8 kpc (Kunder \& Chaboyer 2009). Dudziak et al. (2000) 
estimated the effective temperature $T_{\rm eff}$ = (7.0$\pm$1.0)$\times$10$^{4}$ K and luminosity 
$L_{\ast}$=(5.4$\pm$0.4)$\times$10$^{3}$ $L_{\odot}$ of the central star 
in their P-I model. Guided by their $T_{\rm eff}$ and $L_{\ast}$, we used a series of theoretical atmosphere models with a range of values of 
$T_{\rm eff}$ to
supply the SED from the central star. We used Thomas Rauch's non-LTE 
theoretical atmosphere model\footnote[10]{see http://astro.uni-tuebingen.de/$\sim$rauch/} 
for halo stars ($[X, Y]$=0, $[Z]$=--1) and we considered the surface gravity 
$\log$\,{\it g}=6.0 and 6.5 cases. We varied $T_{\rm eff}$ and $L_{\star}$ to match the observations. 
For the elemental abundances $N$(X)/$N$(H), we used the observed values listed in
Table \ref{element} as a first guess. For initial N and O abundances, 
we adopted the values derived from the CELs. Note that line fluxes of ORLs from N and O calculated by the models 
would be underestimated. We did not consider the F, P, and Kr abundances 
because {\sc Cloudy} does not calculate [F\,{\sc ii}]$\lambda$4790, [Kr\,{\sc iii}]$\lambda$6826, 
[P\,{\sc ii}]$\lambda$7875 and also because very small abundances of these elements 
are not effective in cooling. For F, P, and Kr and the other unobserved elements, we therefore fixed [X/H]=--0.6. 
We fixed the outer nebular shell $R_{\rm out}$ at 0.26$''$, which corresponds to the FWHM of the 
$HST$ image (Figure \ref{image}). We adopted a $R^{-2}$ hydrogen density ($N_{\rm H}$) 
profile. We also varied $N$(X)/$N$(H), $N_{\rm H}$ 
at the inner radius, and the inner radius $R_{\rm in}$ over a small range of values to match the observed line 
fluxes detected in the FORS2, MMIRS, and IRS and 2MASS $JHKs$ bands and 
our interesting mid-infrared band. 
IRS B is the integrated flux between 17 and 23 $\mu$m and IRS C is between 27 and 33 $\mu$m.

Dust grains co-exist along with the gas in the nebula of Hen2-436. As discussed in Sections 2.3 and 3.4, Hen2-436 might have PAHs. 
Therefore, we considered two dust composition models; i) 
amorphous carbon (am C, hereafter) only (model A hereafter) and ii) am C + PAH grains (model B, hereafter).
The optical constants were taken from Rouleau \& Martin (1991) for amorphous carbon and taken from
Desert et al. (1990), Schutte et al. (1993), Geballe (1989), and Bregman et al. (1989) 
for PAHs. In the {\sc Cloudy} model we assumed that the gas and dust co-exist in the same sized ionized nebula, so that 
we consider the warm dust ($>$100 K) only. We adopted a standard MRN $a^{-3.5}$ 
distribution (Mathis, Rumpl \& Nordsieck 1977) with $a_{\rm min}$=0.001 
$\mu$m and $a_{\rm max}$=0.25 $\mu$m for amorphous carbon. For PAHs
we adopted an $a^{-4}$ size distribution with $a_{\rm min}$=0.00043 $\mu$m and $a_{\rm
max}$=0.0011 $\mu$m. We adopted an $R^{-2}$ dust density distribution. The dust size distribution 
and dust (and hydrogen) density distribution law are the same as those used in the Sgr dwarf galaxy PN BoBn1 (Otsuka et al. 2010). 
In BoBn1, Otsuka et al. (2010) used an am C + PAH grain model and estimated a dust mass of 5.78(--6) $M_{\odot}$ 
and temperature of 80--180 K.

In Table \ref{cloudy}, we compare the predicted relative line intensities with observed values, where $I$({\hb}) 
is 100. The $\log\,g$=6.0 models give a better fit to the observations. 
Columns 3, 4, and 5 list the observed values and the values predicted by models A and B with $\log\,g$=6.0, 
respectively. Column 6 lists the observed values by Walsh et al. (1997) as a reference. 
In general, the predicted line and wide bands fluxes (except for {\oii}$\lambda$3726/29 and 2MASS $Ks$) 
agree with the observations within 30$\%$ for models A and B. The large discrepancy between the observed 
{\oii}$\lambda$3726/29 line fluxes and the model might be due to the flux calibration uncertainty 
around 3700 {\AA} and the adopted monotonically decreasing $R^{-2}$ density profile. Since the models did not 
include a density jump around the ionization front which sometimes weakens the line intensity by collisional 
de-excitation, the models would predict larger {\oii}$\lambda$3726/29 fluxes than 
the observations. The large discrepancy between the 2MASS $Ks$ and the models might be due to the contribution from the H$_{2}$ line.

\begin{figure}
\epsscale{1.1}
\plotone{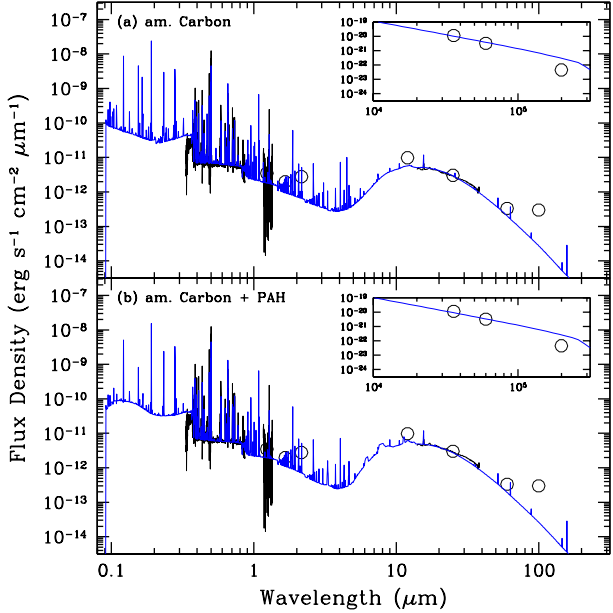}
\caption{The predicted SED from the P-I modeling (blue lines) by {\sc Cloudy}.
The observed data are indicated by the black lines or circles. In the inner small boxes we plot the radio data and the predicted SED. 
({\it upper panel}) The predicted SED by {\sc Cloudy} considering amorphous carbon only. ({\it lower panel}) The predicted SED considering 
amorphous carbon and PAH grains (see text). \label{sed}}
\end{figure}

\begin{table}
\centering
\caption{The predicted relative line fluxes by the {\sc Cloudy} models.\label{cloudy}}
\begin{tabular}{lcrrr|r}
\hline\hline
Ion &$\lambda$&$I$($\lambda$)$_{\rm obs}$&$I$($\lambda$)$_{\rm cloudy}$&$I$($\lambda$)$_{\rm cloudy}$&Walsh et al.\\
    &({\AA}/$\mu$m)&&Model A&Model B&(1997)\\
\hline 
$[$O~{\sc ii}$]$   &3727    &9.25  &27.80&25.60&10.86\\
$[$Ne~{\sc iii}$]$ &3869    &53.57 &61.64&56.87&57.70\\
$[$S~{\sc ii}$]$   &4068    &2.07  &2.95&3.12&2.78\\
$[$S~{\sc ii}$]$   &4076    &0.81  &0.95&1.00&0.98\\
C~{\sc ii}         &4267    &1.10  &1.19&1.31&0.97\\
O~{\sc ii}         &4294    &0.04  &0.02&0.03&\nodata\\
$[$O~{\sc iii}$]$  &4363    &13.90 &12.37&10.84&14.31\\
He~{\sc i}         &4471    &5.91  &6.99&6.53&6.11\\
$[$Ar~{\sc iv}$]$  &4740 &0.73  &0.99&0.82&0.74\\
$[$O~{\sc iii}$]$  &4931 &0.15&0.11&0.11&\nodata\\
$[$O~{\sc iii}$]$  &4959 &265.07&259.39&262.20&271.64\\
$[$O~{\sc iii}$]$  &5007 &792.07&780.77&789.22&790.83\\
$[$Ar~{\sc iii}$]$ &5192 &0.06  &0.11&0.13&\nodata\\
$[$Fe~{\sc iii}$]$ &5271 &0.08  &0.24&0.15&\nodata\\
$[$Cl~{\sc iii}$]$ &5518 &0.06  &0.04&0.05&0.21\\
$[$Cl~{\sc iii}$]$ &5538 &0.16  &0.15&0.17&0.13\\
$[$N~{\sc ii}$]$   &5755 &0.72  &0.70&0.73&1.10     \\
He~{\sc i}         &5876 &19.56 &22.36&20.73&18.21\\
$[$S~{\sc iii}$]$  &6312 &1.63  &1.06&1.14&2.47\\
$[$N~{\sc ii}$]$   &6548 &3.59  &3.59&3.80&3.54\\
$[$N~{\sc ii}$]$   &6583 &10.92 &10.61&11.22&11.43\\
He~{\sc i}     &6678     &4.92  &5.20&4.88&4.77\\
$[$S~{\sc ii}$]$   &6716 &0.49  &0.69&0.66&0.49\\
$[$S~{\sc ii}$]$   &6731 &1.10  &1.42&1.41&1.10\\
$[$Ar~{\sc iii}$]$ &7135 &7.66  &8.03&11.16&6.48\\
$[$Ar~{\sc iv}$]$  &7171 &0.04  &0.04&0.03&\nodata\\
$[$Ar~{\sc iv}$]$  &7263 &0.04  &0.03&0.02&\nodata\\
$[$O~{\sc ii}$]$   &7323 &8.29  &16.84&15.74&6.24\\
$[$O~{\sc ii}$]$   &7332 &6.75  &13.59&12.69&7.03\\
$[$Ar~{\sc iii}$]$ &7751 &1.84  &1.94&2.69&1.22\\
$[$Cl~{\sc iv}$]$  &8047 &0.17  &0.21&0.25&\nodata\\
C~{\sc iii}        &8197 &0.05  &0.13&0.10    &\nodata\\  
He~{\sc i}         &1.28 &1.13  &0.98    &0.93       &\nodata  \\ 
H~{\sc i}          &1.28 &16.18 &15.63&15.61&\nodata\\
$[$Ne~{\sc iii}$]$ &15.55&31.41 &29.87&30.92&\nodata\\
\hline
2MASS $J$          &1.24 &51.90 &49.06  &50.57&     \nodata\\
2MASS $H$          &1.66 &40.00 &27.37  &32.20&     \nodata\\
2MASS $Ks$         &2.16 &98.50 &22.21 &23.81&     \nodata\\
IRS B              &20.0 &2182.05  &2199.71&2044.33&\nodata\\    
IRS C              &30.0 &1004.30   &979.57&922.92&\nodata\\  
\hline
\end{tabular}
\end{table}

\begin{table}
\centering
\caption{The derived properties of the PN central star, ionized nebula, and dust by
 the P-I model. \label{cloudypara}}
\begin{tabular}{@{}lll@{}}
\hline\hline
\multicolumn{3}{c}{Central star}  \\ 
parameters      &Model A&Model B\\
\hline 
$\log,L_{\star}/L_{\odot}$     &3.60&3.57\\
$\log, T_{eff}$ (K) &5.07       &4.93        \\
$\log\,g$ (cm$^{2}$ s$^{-1}$)&6.0&6.0\\
composition     &$[X,Y]$=0, $[Z]=-1$&$[X,Y]$=0, $[Z]=-1$\\
\hline\hline
\multicolumn{3}{c}{Nebula}\\     
parameters      &Model A&Model B\\
\hline 
composition     &He:11.15,C:9.17,N:6.93,&He:11.08,C:9.19,N:7.03,           \\
          &O:8.29,Ne:7.43,S:6.15,       &O:8.33,Ne:7.45,S:6.25,\\
          &Cl:4.33,Ar:5.74,Fe:5.72  &Cl:4.43,Ar:5.89,Fe:5.57\\
          &others:[X/H]=--0.6       &others:[X/H]=--0.6\\
geometry  &spherical                &spherical\\
$R_{\rm in}$/$R_{\rm out}$ ($''$)&0.19/0.29 & 0.18/0.26\\
$\log,N_{\rm H}(R_{\rm in})$ (cm$^{-3}$)&4.75   &4.78\\              
$\log F$({\hb}) &--12.02&--12.02\\
$M_{\rm gas}$ ($M_{\odot}$)&0.05&0.07\\
dust comp.&am. C&am. C + PAHs\\
$T_{\rm dust}$ (K)&100--150&100--200\\ 
$M_{\rm dust}$ ($M_{\odot}$)&2.9(--4)&4.0(--4)\\
$M_{\rm dust}$/$M_{\rm gas}$&5.92(--3)&5.58(--3)\\
\hline

\end{tabular}
\end{table}

\begin{table}
\centering
\caption{Gas and grain carbon mass in Hen2-436. \label{cmass}}
\begin{tabular}{lcc}
\hline\hline
parameters                      &Model A  &Model B\\
\hline
gas C ($M_{\odot}$)        &1.3(--3)&8.6(--4)\\
grain C ($M_{\odot}$)      &2.9(--4)&4.0(--4)\\
gas + grain C ($M_{\odot}$)&1.6(--3)&1.3(--3)\\
$N$(gas + grain C)/$N$(H)$^{\dagger}$     &1.8(--3)&2.3(--3)\\
\hline
\end{tabular}
\tablenotetext{$^{\dagger}$}{number density}
\end{table}

In Table \ref{cloudypara}, we list the derived parameters of the PN central star,
ionized gas nebula, and dust. In Figure \ref{sed} we present the predicted 
SED from {\sc Cloudy} (blue lines). The upper and lower panels are the results of models 
A and B, respectively. In the inner small boxes we plot the radio flux densities (circles) 
from Dudziak et al. (2000) and the predicted SED. Dudziak et al. (2000) measured flux 
densities of 0.62/3.90/4.90 mJy at 1.46, 4.89, 8.40 GHz respectively. 
Although the predicted SED well matches the data in optical to radio wavelengths, 
it cannot explain the $\sim$30 $\mu$m bump and the $IRAS$ 100 $\mu$m data.  
The 30~$\mu$m bump might be from Magnesium Sulfide (MgS) grains, which are sometimes observed in C-rich PNe. 
At the present, since there are no available optical constants 
of MgS at UV wavelengths, we did not consider MgS in the SED models. 
Most of the IRAS 100 $\mu$m flux would be from the surrounding cold ISM because the spatial 
resolving power is $\sim$2$'$ at this band.  Accordingly, our models failed to explain the flux in these bands. 

For Hen2-436, this work is the first successful estimate of the dust mass of 2.9(--4) $M_{\odot}$ 
in model A and 4.0(--4) $M_{\odot}$ in model B and temperatures of 100-150 K in model A and 100--200 K in model B. 
The derived dust-to-gas mass ratios in both models are slightly larger than the typical value in PNe having nebular radius of 
10$^{17}$ cm ($\sim$10$^{-3}$; Pottasch 1984). The value is rather close to the LMC ISM (4$\times$10$^{-3}$; Meixner et al. 2010). 
Lagadec et al. (2009, 2010) assumed a dust-to-gas mass ratio of 5$\times$10$^{-3}$ in radiative transfer modelings 
of C-rich stars in the Sgr dwarf galaxy.

We feel that model B can better explain the physical conditions of gas and dust in Hen2-436, because the line fluxes predicted 
by model B give a better fit to the observations and $T_{\rm eff}$ is not so high as 
to produce strong emission from He~{\sc ii}, which would originate in the stellar wind of Hen2-436. The predicted $T_{\rm eff}$ and $L_{\ast}$ are comparable 
to the values by Dudziak et al. (2000). At this time, there is no UV data of Hen2-436. 
The flux density profile in the UV region and the mid IR flux profile depend upon the central star temperature 
and luminosity as well as the dust composition. 
The predicted SEDs in the optical or longer wavelength regions by Models A and B do not show large differences, 
however a slight difference is found around 0.1--0.2 $\mu$m as can be seen in Figure \ref{sed}. 
This discrepancy could be due to dust composition. Certainly in PNe, the bulk of the heating of the grains will be done by UV photons around 0.1 $\mu$m, 
where the dust absorption coefficient peaks. To further enlarge our understanding, UV and also 3--10 $\mu$m 
spectra would be necessary. These data would enable us to estimate the C abundance using 
C~[{\sc iii}] $\lambda\lambda$1906/09 lines and the CEL C/O ratio and also check for the existence of PAHs.

In general, the elemental abundances of PNe have been estimated based on gas emission lines only. Here, we can estimate 
gas and grain C mass through SED modeling. We can then investigate how much the C abundance increase if we include the contribution of grains.
In Table \ref{cmass}, we present the gas and grain phase C mass in both models. It is remarkable that in model B, which is the best model at the present, 
$\sim$50 $\%$ of the gas C mass exists as grains. If the C abundance is measured using the sum of gas and grain C masses, 
the C abundance ($\log_{10} N$(C)/$N$(H)+12) would be 9.36 dex, which is $\sim$0.17 dex larger than the case of only gas C (9.19 in the case of gas C only, 
see Table \ref{cloudypara}). The grain C abundance might be considerable in PNe.

\subsection{Evolutionary Status}
In Figure \ref{hr} we plot the location of Hen2-436 and two Sgr dwarf galaxy PNe StWr2-21 and Wray16-423 (Zijlstra et al. 2006) 
and the post-AGB He-burning evolutionary tracks with $Z$ = 0.008 by Vassiliadis \& Wood (1994). Since all 
these three Sgr dwarf galaxy PN have WR type central stars (Zijlstra et al. 2006), we assume that the central stars are hydrogen-poor. 
For Hen2-436, we find that $L_{\ast}$ predicted by models A and B has and uncertainty of $\sim$1200 $L_{\odot}$, which is estimated 
from the {\hb} flux and distance determination errors.  The uncertainty of $T_{\rm eff}$ is 
$\sim$10\,000 K. These evolutionary tracks suggest that the progenitors are 1.5--2 $M_{\odot}$ stars, 
which end their lives as white dwarfs with a core mass of $\sim$0.63--0.67 $M_{\odot}$. Among the above three PNe, Hen2-436 seems to be 
the youngest PN. The age of Hen2-436 is estimated from the evolutionary tracks to be $\sim$3000 yr after leaving the AGB phase. 
Gesicki \& Zijlstra (2000) estimated the expansion velocity to be 14 {\kms}. When adopting the value  $R_{out}$ = 0.26$''$ 
(see Table \ref{cloudypara}) and the distance of 24.8 kpc, 
the dynamical age is estimated to be $\sim$2200 yr, which is consistent with the evolutionary age.

\subsection{Dust Mass-Loss Rate}
If most of the observed dust is formed during the last two thermal pulses ($\sim$10\,000 yr, 
Vassiliadis \& Wood 1993 for the case of a 2.0 $M_{\odot}$ initial mass 
and $Z$ = 0.008), the dust mass-loss rate  ($\dot{M}_{\rm dust}$) is estimated to be $<$3.1(--8) $M_{\odot}$ yr$^{-1}$. 
Lagadec et al. (2010) estimated a $\dot{M}_{\rm dust}$ of 0.97--2.52(--8) $M_{\odot}$ yr$^{-1}$ for IRAS18436-2849 in the Sgr dwarf galaxy. 
IRAS18436-2849 shows similar metallicity and luminosity to Hen2-436. For a C-rich metal-poor star ([Fe/H]$\sim$--1) 
IRAS12560+1656 in the Sgr dwarf galaxy stream, Groenewegen et al. (1997) estimated $\dot{M}_{\rm dust}$ to be 1.9(--9) $M_{\odot}$ yr$^{-1}$. 
Our estimated value is comparable to their values. Adopting a dust-to-gas mass ratio of 5.58(--3), the mass-loss rate 
$\dot{M}$ during  the last two thermal pulses is estimated to be $<$5.5(--6) $M_{\odot}$ yr$^{-1}$, which is very close 
to the predicted mass-loss rate during the last two thermal pulses (Vassiliadis \& Wood 1993 for the same case above).

\begin{figure}
\epsscale{1.1}
\plotone{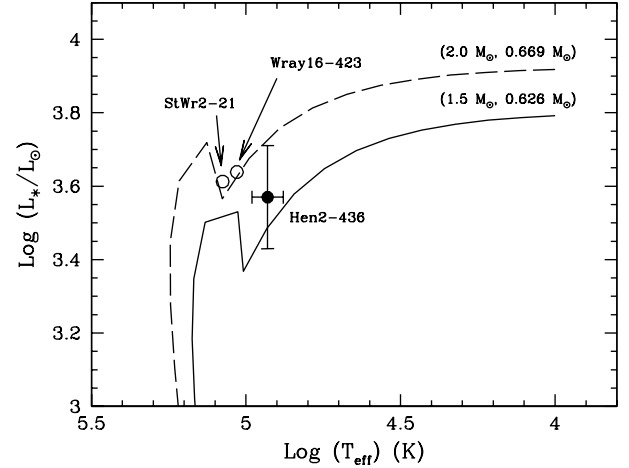}
\caption{The evolutionary tracks for He-burning stars with $Z$ = 0.008 and the positions of three Sgr dwarf galaxy PNe.}
\label{hr}
\end{figure}

\begin{table}
\centering
\caption{The observed chemical abundances and the theoretical model.\label{theo}}
\begin{tabular}{@{}ccccccccccl@{}}
\hline\hline
Nebula                       &He&C&N&O&F&Ne&P&S&Ref\\
\hline
Hen2-436                    &11.09&9.08&7.38&8.35&5.69   &7.62&5.72&6.83&(1)\\
StWr2-21                    &11.00&9.00&7.88&8.53&\nodata&7.54&\nodata&6.93&(2)\\
Wray16-423                  &11.03&8.86&7.68&8.33&\nodata&7.55&\nodata&6.67&(2)\\
\hline
Model\\
\hline
2.25 $M_{\odot}$ ($Z=0.008$)&11.01&9.07&8.00&8.54&5.00&7.95&5.18&6.85&(3)\\
\hline
\end{tabular}
\tablerefs{(1) This work; (2) Zijlstra et al. (2006); (3) Karakas (2010)}
\end{table}

\subsection{Comparison between the Observed Chemical Abundances and the Theoretical Model}
In Table \ref{theo}, we present the observed elemental abundances of Hen2-436, StWr2-21, and Wray16-423 (Zijlstra et al. 2006) 
and predicted values from the theoretical models of Karakas (2010) for 2.25 $M_{\odot}$ stars with initial $Z$=0.008. 
The abundances of Sgr dwarf galaxy PNe are measured using gas emission
lines only and their C abundances are derived from the ORL C lines.
abundances from the models are the values at the end of the AGB. Karakas (2010) predicted that such stars end as white dwarfs with a core 
mass of 0.652 $M_{\odot}$, which is comparable to the core mass of Hen2-436 estimated from the evolutionary tracks. In the model for 
2.25 $M_{\odot}$ stars with $Z$=0.008, a partial mixing zone, which produces a $^{13}$C pocket during the interpulse 
period and releases free neutrons, is not included and the mass-loss law during the AGB phase by Vassiliadis \& Wood (1993) is adopted.
For the models with $Z$=0.008, Karakas (2010) adopted LMC compositions from Russell \& Dopita (1992). 
The accuracy of the predicted abundances by the models is on the order of 0.3 dex (A. Karakas in private communication). For the observed N 
and O abundances of Hen2-436, we adopted the values derived from the CELs. 

In general the observed chemical abundances of the above three PNe agree well with the model of Karakas (2010) 
within estimation errors, which suggest that all three of these PNe have evolved from $\approx$2.25 $M_{\odot}$ single stars 
with initial LMC chemical compositions. The three Sgr dwarf galaxy PNe would have experienced evolution similar to LMC PNe, which 
is a remarkable finding. Some of the $^{14}$N in Hen2-436 might have been converted into $^{22}$Ne by the double 
$\alpha$ particle capture process. The overabundances of $^{19}$F and $^{31}$P in Hen2-436 imply that an extensive 
partial mixing zone was formed and the extra neutrons were released in the He-rich intershell.

\section{Conclusion}
We estimated elemental abundances in the Sgr dwarf galaxy PN Hen2-436 based on the archived ESO/VLT 
FORS2 and $Spitzer$/IRS spectra. We detected candidates of [F\,{\sc
ii}]$\lambda$4790, [Kr\,{\sc iii}]$\lambda$6826, and [P\,{\sc
ii}]$\lambda$7875 for the first time, which indicates that these elements 
are largely enhanced. We found a co-relation between 
C and F, P, and Kr abundances among PNe and C-rich stars. The detections of 
F, P and Kr in Hen2-436 support that F, P, and Kr together with C are certainly 
synthesized in the same layer and brought to the stellar surface by the third dredge-up. 
We detected some N~{\sc ii} and O~{\sc ii} ORLs 
and derived the ionic abundances from these lines. The discrepancy between O ORL 
and CEL abundances is $>$1 dex. We constructed a SED model considering dust and  
estimated the initial mass of the progenitor to be $\sim$1.5-2.0 $M_{\odot}$ with $Z$=0.008 
and the age to be $\sim$3000 yr after the AGB phase. Hen2-436 shares its evolutionary status 
with the Sgr dwarf galaxy PNe StWr2-21 and Wray16-423. The observed elemental abundances of 
the three Sgr dwarf galaxy PNe could be explained by a theoretical nucleosynthesis model with initial 
mass 2.25 $M_{\odot}$, $Z$=0.008, and LMC chemical abundances. The SED model predicted that 
$>$2.9(--4) $M_{\odot}$ of carbon dust co-exists in the ionized nebula. Hen2-436 seems to have experienced 
evolution similar to LMC PNe.  Based on the assumption that 
most of the observed dust is formed during the last two thermal pulses and the dust-to-gas mass ratio is 
5.58(--3), the dust mass-loss rate and the total mass-loss rate are $<$3.1(--8) $M_{\odot}$ yr$^{-1}$ 
and $<$5.5(--6) $M_{\odot}$ yr$^{-1}$, respectively. Our estimated dust mass-loss rate is comparable to 
a similar metallicity and luminosity Sgr dwarf galaxy AGB star.

\acknowledgements 
The authors express their thanks to Amanda Karakas for a fruitful discussion of AGB nucleosynthesis and a critical reading of the manuscript. 
They wish to thank the anonymous referee for valuable comments. 
M.O. and M.M. acknowledge funding support from STScI GO-1129.01-A and NASA NAO-50-12595. M.O. 
acknowledges funding support form STScI DDRF D0101.90128. M.M. appreciates support from Harvard-Smithsonian 
Center for Astrophysics during this work. 
S.H. acknowledges the support by Basic Science Research Program through the National Research Foundation 
of Korea funded by the Ministry of Education, Science and Technology (NRF-2010-0011454).
We thank to the Magellan Telescope staffs for supporting MMIRS observations.
This work is in part based on ESO archive data obtained by ESO Telescopes at the Paranal Observatory. 
This work is in part based on archival data obtained with the Spitzer Space Telescope, which is operated by the 
Jet Propulsion Laboratory, California Institute of Technology under a contract with NASA. Support for this 
work was provided by an award issued by JPL/Caltech. This work in in part based on $HST$ archive data 
downloaded from the Canadian Astronomy Data Centre.



\begin{thebibliography}{}
\bibitem[]{} Abia, C., et al.\ 2010, \apjl, 715, L94 
\bibitem[]{} Acker, A., \& Neiner, C.\ 2003, \aap, 403, 659 
\bibitem[]{} Appenzeller, I., et al.\ 1998, The Messenger, 94, 1 
\bibitem[]{} Baker, J.~G., \& Menzel, D.~H.\ 1938, \apj, 88, 52 
\bibitem[]{} Benjamin, R.~A., Skillman, E.~D.,  \& Smits, D.~P.\ 1999, \apj, 514, 307 
\bibitem[]{} Bernard-Salas, J., Peeters, E., Sloan, G.~C., Gutenkunst, S., Matsuura, M., Tielens, A.~G.~G.~M., Zijlstra, A.~A., \& Houck, J.~R.\ 2009, \apj, 699, 1541 
\bibitem[]{} Bi{\'e}mont, E., \& Hansen, J.~E.\ 1986, \physscr, 34, 116 
\bibitem[]{} Bregman, J.~D., Allamandola, L.~J., Witteborn, F.~C., Tielens, A.~G.~G.~M.,  \& Geballe, T.~R., 1989, ApJ, 344, 791
\bibitem[]{} Busso, M., Gallino, R., \& Wasserburg, G.~J.\ 1999, \araa, 37, 239 
\bibitem[]{} Cahn, J.~H., Kaler, J.~B.,  \& Stanghellini, L.\ 1992, \aaps, 94, 399 
\bibitem[]{} Cardelli, J.~A., Clayton, G.~C.,  \& Mathis, J.~S.\ 1989, \apj, 345, 245
\bibitem[]{} Cohen, M., \& Barlow, M. J. 2005, MNRAS, 362, 1199
\bibitem[]{} De Marco, O., \& Barlow, M.~J.\ 2001, \apss, 275, 53 
\bibitem[]{} Desert, F.-X., Boulanger, F., \& Puget, J.L. 1990, A\&A, 237, 215
\bibitem[]{} Dudziak, G., P{\'e}quignot, D., Zijlstra, A.~A., \& Walsh, J.~R.\ 2000, \aap, 363, 717 
\bibitem[]{} Ferland, G.~J.\ 2004, Bulletin of the American Astronomical Society, 36, 1574 
\bibitem[]{} Geballe, T.~R., 1989, ApJ, 344, 791
\bibitem[]{} Gesicki, K., \& Zijlstra, A.~A.\ 2000, \aap, 358, 1058 
\bibitem[]{} Girard, P., K{\"o}ppen, J., \& Acker, A.\ 2007, \aap, 463, 265 
\bibitem[]{} Groenewegen, M.~A.~T., Oudmaijer, R.~D.,  \& Ludwig, H.-G.\ 1997, \mnras, 292, 686 
\bibitem[]{} Herwig, F.\ 2005, \araa, 43, 435 
\bibitem[]{} Houck, J.~R., et al.\ 2004, SPIE, 5487, 62 
\bibitem[]{} Hyung, S., Aller, L.~H., Feibelman, W.~A., \& Lee, S.-J.\ 2001, \apj, 563, 889 
\bibitem[]{} Hyung, S., \& Aller, L.~H.\ 1997, \mnras, 292, 71 
\bibitem[]{} Hyung, S., Aller, L.~H., \& Feibelman, W.~A.\ 1994, \mnras, 269, 975 
\bibitem[]{} Karakas, A.~I.\ 2010, \mnras, 403, 1413 
\bibitem[]{} Karakas, A. I., van Raai, M. A., Lugaro, M., Sterling, N. C., \& Dinerstein, H. L. 2009, ApJ, 690, 1130
\bibitem[]{} Karakas, A., \& Lattanzio, J.~C.\ 2007, PASA, 24, 103 
\bibitem[]{} Kingdon, J., \& Ferland, G.~J.\ 1995, \apj, 442, 714 
\bibitem[]{} Kniazev, A.~Y., et al. 2008, \mnras, 388, 1667
\bibitem[]{} Kunder, A., \& Chaboyer, B.\ 2009, \aj, 137, 4478
\bibitem[]{} Lagadec, E., Zijlstra, A.~A., Mauron, N., Fuller, G., Josselin, E., Sloan, G.~C., \& Riggs, A.~J.~E.\ 2010, \mnras, 403, 1331 
\bibitem[]{} Lagadec, E., et al.\ 2009, \mnras, 396, 598 
\bibitem[]{} Liu, X.-W., Luo, S.-G., Barlow, M. J., Danziger, I. J.,  \& Storey, P.~J. 2001, {\mnras}, 327, 141
\bibitem[]{} Liu, X.-W., Storey, P. J., Barlow, M. J.,  \& Clegg, R. E. S. 1995, {\mnras}, 272, 369
\bibitem[]{} Liu, X.-W., Storey, P.~J., Barlow, M.~J., Danziger, I.~J., Cohen, M.,  \& Bryce, M.\ 2000, \mnras, 312, 585
\bibitem[]{} Liu, Y., Liu, X.-W., Barlow, M.~J., \& Luo, S.-G.\ 2004, \mnras, 353, 1251 
\bibitem[]{} Lugaro, M., Ugalde, C., Karakas, A.~I., G{\"o}rres, J., Wiescher, M., Lattanzio, J.~C., \& Cannon, R.~C.\ 2004, \apj, 615, 934 
\bibitem[]{} Lodders, K.\ 2003, \apj, 591, 1220 
\bibitem[]{} Marcolino, W.~L.~F., Hillier, D.~J., de Araujo, F.~X., \& Pereira, C.~B.\ 2007, \apj, 654, 1068 
\bibitem[]{} Mathis, J.~S., Rumpl, W., \& Nordsieck, K.~H.\ 1977, \apj, 217, 425 
\bibitem[]{} McLeod, B.~A., Fabricant, D., Geary, J., Martini, P., Nystrom, G., Elston, R., Eikenberry, S.~S., \& Epps, H.\ 2004, \procspie, 5492, 1306 
\bibitem[]{} Meixner, M., et al.\ 2010, \aap, 518, L71 
\bibitem[]{} Mendoza, C., \& Zeippen, C.~J.\ 1982, \mnras, 199, 1025 
\bibitem[]{} Otsuka, M., Tajitsu, A., Hyung, S., \& Izumiura, H. 2010, \apj, 723, 658 
\bibitem[]{} Otsuka, M., Izumiura, H., Tajitsu, A.,  \& Hyung, S.\ 2008, \apjl, 682, L105 
\bibitem[]{} Pottasch, S.~R., Bernard-Salas, J., \& Roellig, T.~L.\ 2009, \aap, 499, 249 
\bibitem[]{} Pottasch, S.~R., Bernard-Salas, J., \& Roellig, T.~L.\ 2008, \aap, 481, 393 
\bibitem[]{} Pottasch, S.~R., Bernard-Salas, J., Beintema, D.~A., \& Feibelman, W.~A.\ 2003, \aap, 409, 599 
\bibitem[]{} Pottasch, S.~R.\ 1984, {\it ``Planetary Nebulae''}, Astrophysics and Space Science Library, 107
\bibitem[]{} Rouleau, F., \& Martin, P.G. 1991, \apj, 377, 526
\bibitem[]{} Ramos-Larios, G., Phillips, J.~P., \& Cuesta, L.~C.\ 2010, \mnras, 1733 
\bibitem[]{} Rudy, R.~J., Rossano, G.~S., Erwin, P., \& Puetter, R.~C.\ 1991, \apj, 368, 468 
\bibitem[]{} Russell, S. C., \& Dopita, M.~A. 1992, ApJ, 384, 508
\bibitem[]{} Schlegel, D.~J., Finkbeiner, D.~P., \& Davis, M.\ 1998, \apj, 500, 525 
\bibitem[]{} Schutte, W.~A., Tielens, A.G.G.M., \& Allamandolla, L.J., 1993, ApJ, 415, 397
\bibitem[]{} Seaton, M.~J.\ 1979, \mnras, 187, 73P 
\bibitem[]{} Schoning, T.\ 1997, \aaps, 122, 277
\bibitem[]{} Sterling, N.~C., et al.\ 2009, Publications of the Astronomical Society of Australia, 26, 339 
\bibitem[]{} Sterling, N.~C., \& Dinerstein, H.~L.\ 2008, \apjs, 174, 158 
\bibitem[]{} Storey, P.~J.,  \& Hummer, D.~G.\ 1995, \mnras, 272, 41 
\bibitem[]{} Surendiranath, R., \& Pottasch, S.~R.\ 2008, \aap, 483, 519 
\bibitem[]{} Tayal, S.~S.\ 2004, \apjs, 150, 465 
\bibitem[]{} Tsamis, Y.~G., Barlow, M.~J., Liu, X.-W., Storey, P.~J., \& Danziger, I.~J.\ 2004, \mnras, 353, 953 
\bibitem[]{} Vassiliadis, E.,  \& Wood, P.~R.\ 1993, \apj, 413, 641 
\bibitem[]{} Vassiliadis, E.,  \& Wood, P.~R.\ 1994, \apjs, 92, 125 
\bibitem[]{} Walsh, J.~R., Dudziak, G., Minniti, D., \& Zijlstra, A.~A.\ 1997, \apj, 487, 651 
\bibitem[]{} Wang, W., \& Liu, X.-W.\ 2007, \mnras, 381, 669 
\bibitem[]{} Wiese, W. L., Fuhr, J. R.,  \& Deters, T. M. 1996, J.Phys.Chem.Ref.Data,Monograph No.7, {\it ``Atomic Transition Probabilities of Carbon, Nitrogen and Oxygen''}, American Chemical Society, Washington,DC, and American Institute of Physics, New York
\bibitem[]{} Wesson, R., Liu, X.-W., \& Barlow, M.~J.\ 2005, \mnras, 362, 424 
\bibitem[]{} Wesson, R., Liu, X.-W., \& Barlow, M.~J.\ 2003, \mnras, 340, 253 
\bibitem[]{} Werner, K., \& Herwig, F.\ 2006, \pasp, 118, 183 
\bibitem[]{} Zhang, Y., \& Liu, X.-W.\ 2005, \apjl, 631, L61 
\bibitem[]{} Zijlstra, A.~A., Gesicki, K., Walsh, J.~R., P{\'e}quignot, D., van Hoof, P.~A.~M.,  \& Minniti, D.\ 2006, \mnras, 369, 875 

\end{thebibliography}
\end{document}